\documentclass[prd,preprintnumbers,showkeys,nofootinbib,12pt]{article}
\usepackage{epsfig}

\usepackage{bm, graphicx,amsmath,amssymb,epsf}
\usepackage{appendix}

\usepackage{verbatim}

\usepackage{setspace}
\usepackage[active]{srcltx}

\usepackage{subfigure}

\usepackage{graphicx}
\newcommand{\beq}{\begin{equation}}
\newcommand{\eeq}{\end{equation}}
\newcommand{\beqn}{\begin{eqnarray}}
\newcommand{\eeqn}{\end{eqnarray}}
\newcommand{\beqs}{\begin{eqnarray*}}
\newcommand{\eeqs}{\end{eqnarray*}}

\begin{document}

\title{\large \bf Collinear and Regge behavior of $2 \to 4$  MHV amplitude in $\mathcal{N}=4$ super Yang-Mills theory}
\author{\large J.~Bartels$^{1}$, L.~N. Lipatov$^{1,2}$ and
A.~Prygarin$^{1}$ \bigskip \\
{\it
$^1$~II. Institute of  Theoretical Physics, Hamburg University, Germany} \\
{\it  $^2$~St. Petersburg Nuclear Physics Institute, Russia}}

\maketitle

\vspace{-9cm}
\begin{flushright}
DESY-11-052
\end{flushright}
\vspace{8cm}

\abstract{
We investigate the collinear and Regge behavior of the  $2\to 4$  MHV amplitude in $\mathcal{N}=4$ super Yang-Mills theory in the BFKL approach.
The expression for the remainder function in the collinear kinematics proposed by Alday, Gaiotto, Maldacena, Sever and Vieira is analytically continued to the Mandelstam region. The result of the continuation in the Regge kinematics shows an agreement with the BFKL approach  up to to five-loop level.  We present the Regge theory interpretation of the obtained results and discuss some issues related to a possible non-multiplicative renormalization of the remainder function in the collinear limit. }

\newpage



\section{Introduction}
The recent developments in the study of the Maximally Helicity Violating~(MHV) amplitudes in $\mathcal{N}=4$ super  Yang-Mills theory encourage us to apply a well studied Balitsky-Fadin-Kuraev-Lipatov~(BFKL) approach to test some analytic results available on the market.  The study of the MHV amplitudes is traced back to the paper of Parke and Taylor~\cite{Parke:1986gb}, who showed that a tree-level gluon scattering amplitude significantly simplifies for a definite helicity configuration of the external particles. The
simplicity of the tree MHV amplitudes suggested that they  could have  some nice properties also at the  quantum level.   This idea led to a formulation of the Anastasiou-Bern-Dixon-Kosower~(ABDK)~\cite{Anastasiou:2003kj}  and later to the Bern-Dixon-Smirnov~(BDS)~\cite{BDS} all-loop formula for multi-leg MHV amplitudes in $\mathcal{N}=4$  SYM.
The BDS ansatz was tested in the regimes of the strong coupling by Alday and Maldacena~\cite{Alday:2007he} and the weak coupling by two of the authors in collaboration with Sabio Vera~\cite{BLS1}. Both of the studies showed some inconsistency of the BDS formula for  a number of the external gluons being larger than five.  At strong coupling  the multi-leg MHV amplitude was considered in the limit of the very large number of external legs using the minimal surface approach~\cite{Alday:2007hr}. At weak coupling, the analytic structure of the BDS amplitude was studied at two loops for four, five and six external gluons in the multi-Regge kinematics~\cite{BLS1}. The BDS amplitude with four and five external gluons were shown to be compatible with the dispersive representation in the Regge kinematics, while  the six gluon BDS amplitude at two loops could not match a form expected from the Regge theory.  
This deficiency becomes especially clear if we consider a physical kinematic region, where some of the energies are negative (this region has been named Mandestam region).
It was argued~\cite{BLS1,BLS2} that the BDS amplitude should be corrected  starting at two loops and six external gluons due to  the fact that it does not account properly for the so-called Regge or Mandelstam cuts in the complex angular momenta plane.  The two loop correction to the six gluon  BDS amplitude was calculated  in the multi-Regge kinematics by two of the authors in collaboration with Sabio Vera~\cite{BLS2} using the Balitsky-Fadin-Kuraev-Lipatov~(BFKL) approach~\cite{BFKL}.

On the other hand recent studies showed an intimate relation between expectation value of polygon Wilson loops and scattering amplitudes in $\mathcal{N}=4$ SYM. It was assumed~\cite{Alday:2007hr} that the BDS formula can be corrected by a multiplicative function named \emph{the remainder function}, which depends only on conformal invariants~(anharmonic ratios) in the dual momenta space~\cite{Drummond:2007au,Drummond:2008vq}. The remainder function for the six-gluon MHV amplitude was calculated by Drummond,  Henn,  Korchemsky and Sokatchev~\cite{hexagon}  and presented in terms of rather complicated four-fold integrals, which were simplified in the quasi-multi-Regge kinematics by Del Duca, Duhr and  Smirnov~\cite{DelDuca:2009au,DelDuca:2010zg}  and expressed in terms of generalized Goncharov polylogarithmic functions of  three  dual conformal cross ratios.  Their   result was greatly simplified by Goncharov, Spradlin, Vergu and Volovich~(GSVV)~\cite{GSVV} using  the theory of motives  and was compactly written in terms of only classical polylogarithms.  The analytic continuation of the GSVV remainder function to the Mandelstam region in the multi-Regge kinematics was performed by two of the authors~\cite{LP1,LP2} reproducing the leading logarithmic prediction of ref.~\cite{BLS2}. It also confirmed~\cite{BLS1,LP2} the validity of the dispersion-like relations for the remainder function in the multi-Regge kinematics found by one of the authors~\cite{LipDisp} for the $2 \to 4$ and $3 \to 3$ scattering amplitudes.
 The six-particle MHV amplitude  at the strong coupling was also investigated by one of the authors in collaboration with Kotanski and Schomerus~\cite{Bartels:2010ej} in the Mandelstam region in the multi-Regge kinematics. The analysis of  the analytic properties of the  system of $Y$-equations allowed to extract the leading asymptotics, which is related to the Pomeron intercept at strong coupling. 

Besides the multi-Regge regime, the remainder function was also considered in the so-called \emph{collinear kinematics}, where two or more external gluons become collinear. In this kinematics the remainder function vanishes, but subleading corrections can provide some information on anomalous dimensions of composite operators in the Operator Product Expansion~(OPE) of the polygonal Wilson loops.     The OPE analysis suggested  by Alday, Gaiotto, Maldacena, Sever and Vieira~(AGMSV)~\cite{OPEmalda}  allowed to make prediction for the collinear behavior of the remainder function at strong and weak coupling in the Euclidean kinematics.
This analysis was extended by Gaiotto, Maldacena, Sever and Vieira~\cite{Gaiotto:2011dt} to reproduce the full two remainder function of the six-particle MHV amplitude. 
It should, however, be kept in mind that this OPE expansion might be quite different from the usual short distance of light cone expansions of color singlet operators. Strictly speaking, in the present case we are dealing with planar amplitudes and all exchange channels are in  adjoint  color states; furthermore,
there could be a non-multiplicative renormalization, i.e.  one can have several operators with different anomalous dimensions.  

In the present study we investigate the AGMSV expression for the remainder function for the six-gluon MHV amplitude at weak coupling and compare it with the BFKL predictions in the double-logarithmic approximation.   We perform analytic continuation of  the $2 \to 4$ amplitude to the Mandelstam region   and extract the leading logarithmic terms in the multi-Regge kinematics   reproducing  the BFKL result up to five loops. 
In order to find this agreement we split the anomalous dimension given in the  AGMSV  formula into two pieces, each of them having poles only in one semiplane. We find that all the known  BFKL contributions come only from one of these two contributions. This agrees with the Regge theory expectation to have a clear separation between  the negative and positive poles for the $s$-channel discontinuities of the remainder function, suggesting a sum of  two exponentiations of the anomalous dimensions, which becomes important already at three loops. The proposed alternative exponentiation  agrees with the Regge theory analysis and coincides with the AGMSV expression at two loops.
The ambiguity between the  two exponentiations can be resolved by taking into account next-to-leading corrections to the eigenvalue of the BFKL Kernel in the adjoint representation, which are currently not available and will be calculated  in the near future.

The content of the paper is presented as follows. In the first section we  overview  the BFKL analysis applied to the $2 \to 4$ scattering MHV amplitude in the multi-Regge kinematics. The  section \ref{sec:collregge} is devoted to the collinear behavior of the remainder function in the Mandelstam region, where we calculate the all-loop expression in the collinear and multi-Regge kinematics with double logarithmic accuracy.  Then we present  details of the analytic continuation of the AGMSV remainder function to the Mandelstam channel and comparison with the BFKL approach  up to five loops. In the section \ref{sec:reggeinter} we consider the interpretation of the obtained result from the point of view of the Regge theory and propose an  alternative exponentiation for the anomalous dimension. The main results are discussed in the last section. Some detailed calculations are presented in the appendices.

\section{Regge limit}\label{sec:regge}

\begin{figure}[htbp]
	\begin{center}
		\epsfig{figure=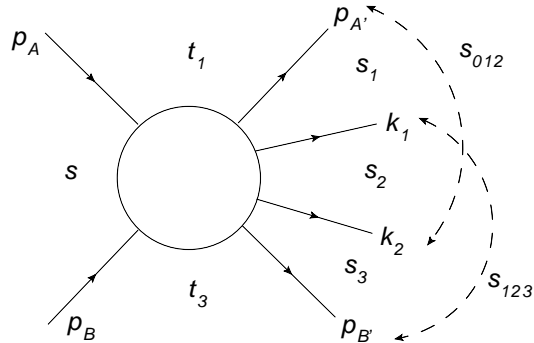,width=70mm}
	\end{center}
	\caption{ The $2 \to 4$ gluon scattering amplitude. }
	\label{fig:2to4ope}
\end{figure}

In this section we discuss the (multi-)~Regge kinematics of the six-gluon scattering MHV amplitude, considered in our previous studies in the regime of  the weak~\cite{BLS1,BLS2,LP1,LP2,BLP1} and strong coupling~\cite{Bartels:2010ej}. The six-gluon amplitude describes to two physical scattering processes, namely to $2 \to 4$ and  $3 \to 3$ scattering. In the present study we  are mainly interested in the $2 \to 4 $ MHV amplitude at weak coupling in the physical channel, where the Mandelstam cuts give a non-vanishing contribution. We call the corresponding channels - the Mandelstam channels. For the purpose of the present discussion it is convenient to introduce the kinematic invariants shown in Fig.~\ref{fig:2to4ope}.

The  invariants  are defined as  $s=(p_A+p_B)^2, \;
s_1=(p_{A'}+k_1)^2, \;  s_{2}=(k_1+k_2)^2, \; s_3=(p_{B'}+k_2)^2,  \;s_{012}=(p_{A'}+k_1+k_2)^2, \;  s_{123}=(p_{B'}+k_1+k_2)^2, \; t_{1}=(p_A-p_{A'})^2, \; t_{2}=(p_A-p_{A'}-k_1)^2 $ and $t_3=(p_{B}-p_{B'})^2$. The dual conformal cross ratios are given by

\beqn\label{crossinv}
u_1=\frac{s s_{2}}{s_{012}\; s_{123}},\; u_2=\frac{s_{1}t_{3}}{s_{012}\; t_2}, \; u_3=\frac{s_{3}t_{1}}{s_{123}  t_2}.
\eeqn

The multi-Regge kinematics, where $s \gg s_{012}\;, s_{123} \gg s_1, \; s_2,\; s_3 \gg |t_1|,\; |t_2|, \; |t_3| $ implies
\beqn\label{multicross}
1-u_1 \to +0,\;\; u_2 \to +0, \;\; u_3 \to +0, \;\; \frac{u_2}{1-u_1} \simeq \mathcal{O}(1), \;\; \frac{u_3}{1-u_1} \simeq \mathcal{O}(1),
\eeqn
which suggests that in this kinematics the convenient variables for the remainder function are $1-u_1$ and the reduced cross ratios  defined by
\beqn\label{redcross}
 \tilde{u}_2=\frac{u_2}{1-u_1}, \;\;\tilde{u}_3=\frac{u_3}{1-u_1}.
\eeqn
In the Regge limit they can be expressed through $s_2$ and the transverse momenta
\beqn\label{redcrossperp}
1-u_1\simeq \frac{(\mathbf{k}_1+\mathbf{k}_2)^2}{s_2}, \;\; \tilde{u}_2\simeq\frac{\mathbf{k}_1^2 \; \mathbf{q}^2_3}{(\mathbf{k}_1+\mathbf{k}_2)^2 \; \mathbf{q}_2^2}, \;\; \tilde{u}_3 \simeq\frac{\mathbf{k}_2^2 \; \mathbf{q}^2_1}{(\mathbf{k}_1+\mathbf{k}_2)^2 \; \mathbf{q}_2^2},
\eeqn
so that the energy $s_2$ dependence of the remainder function is related only to a dependence on $u_1$ and not on $\tilde{u}_2$ and $\tilde{u}_3$. This is not the only choice for expressing  the energy dependence in terms of the  dual cross ratios, but we do not consider other choices  for the sake of   clarity of the presentation.

\begin{figure}[htbp]
	\begin{center}
		\epsfig{figure=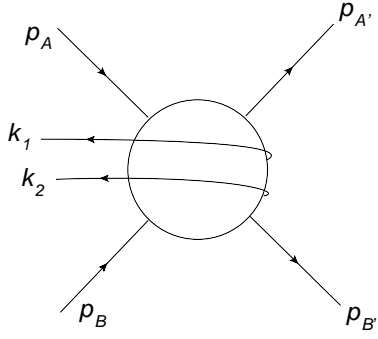,width=50mm}
	\end{center}
	\caption{ The Mandelstam channel of the $2 \to 4$ gluon planar scattering amplitude. }
	\label{fig:2to4opeINV}
\end{figure}

In the ``Euclidean'' kinematics~( $s,\;s_2<0$)  the remainder function vanishes as it follows from the analysis presented in refs.~\cite{BLS1,BLS2}. However, these studies also show that this is not the case in a slightly different physical region, where one  or more dual conformal cross ratios possess a phase.  This happens when some energy invariants change the sign. In the present paper we consider one of such  regions of the  $2 \to 4$ scattering amplitude having
\beqn\label{u1cont}
u_1=|u_1| e^{-i2\pi},
\eeqn
together with $u_2$ and  $u_3$  held fixed and positive.
This corresponds to a physical region~(the Mandelstam channel), where
\beqn\label{mandels}
s\; ,s_2> 0; \;\;\;s_{1},\;s_{3},\;s_{012},\;s_{123}<0
\eeqn
as illustrated in Fig.~\ref{fig:2to4opeINV}. It is worth emphasizing that the scattering amplitude in Fig.~\ref{fig:2to4opeINV} is still planar, but the produced particles have reversed momenta $k_1$ and $k_2$ with negative energy components.

In the Mandelstam channel  the remainder function grows with energy and was first calculated using the BFKL approach by two of the authors in collaboration with A.~Sabio Vera in ref.~\cite{BLS2}. The BFKL approach, based   on the analyticity and  unitarity was developed more than thirty years ago~\cite{BFKL}. In this approach  one sums the contributions from the Feynman diagrams, which are enhanced by the logarithms of the energy~($1-u_1\simeq (\mathbf{k}_1+\mathbf{k}_2)^2/s_2$ in our case). The Leading Logarithmic Approximation~(LLA) allows to write an integral representation of the remainder function $R^{LLA}_{BFKL}$ to any order of the parameter $g^2  \ln s_2 $. The amplitude  in this Mandelstam channel  is given by~\cite{BLS2}
\beq
M_{2\rightarrow 4}=M^{BDS}_{2\rightarrow 4}\,R^{}_{BFKL}=M^{BDS}_{2\rightarrow 4}\,(1+i\Delta _{2\rightarrow 4}),
\label{corLLA}
\eeq
where $M^{BDS}_{2\rightarrow 4}$ is the BDS expression~\cite{BDS} and the correction $\Delta _{2\rightarrow 4}$ was calculated
in all orders with a leading logarithmic accuracy using the solution to the  BFKL eigenvalue in the adjoint representation. The all-order LLA expression  for $\Delta_{2\to 4}$ reads
\beqn\label{LLA}
&& \Delta^{LLA} _{2\rightarrow 4}=\frac{a}{2}\, \sum _{n=-\infty}^\infty (-1)^n
\int _{-\infty}^\infty \frac{d\nu }{\nu ^2+\frac{n^2}{4}}\,
\left(\frac{q_3^*k^*_1}{k^*_2q_1^*}\right)^{i\nu -\frac{n}{2}}\,
\left(\frac{q_3k_1}{k_2q_1}\right)^{i\nu +\frac{n}{2}}\,
\left(s_2^ {\omega (\nu , n)}-1\right)\,\\
&&
\simeq \frac{a}{2}\, \sum _{n=-\infty}^\infty (-1)^n
\int _{-\infty}^\infty \frac{d\nu }{\nu ^2+\frac{n^2}{4}}\,
\left(w^*\right)^{i\nu -\frac{n}{2}}\,
\left(w\right)^{i\nu +\frac{n}{2}}\,
\left((1-u_1)^ {-\omega (\nu , n)}-1\right) \nonumber
\eeqn
Here $k_1,k_2$ are complex transverse components of the gluon momenta,
$q_1,q_2,q_3$ are the corresponding  momenta of reggeons in the
crossing channels. It is convenient to define holomorphic and antiholomorphic variables  in the transverse space as
\beqn
w =\frac{q_3k_1}{k_2q_1}, \;\; w^* =\frac{q^*_3k^*_1}{k^*_2q^*_1}
\eeqn
related to the reduced  cross ratios of (\ref{redcross}) by
\beqn\label{wdef}
 |w|^2=\frac{\tilde{u}_2}{\tilde{u}_3}=\frac{u_2}{u_3}, \;\; w=|w| e^{i(\phi_2-\phi_3)},\;\; \cos (\phi_2-\phi_3)=\frac{1-\tilde{u}_2-\tilde{u}_3}{2 \sqrt{\tilde{u}_2 \tilde{u}_3}}=\frac{1-u_1-u_2-u_3}{2 \sqrt{u_2 u_3}}.
\eeqn
 The energy behavior of the remainder function is determined by the reggeon intercept
\beq\label{eigen}
\omega (\nu , n)=-a E_{\nu,n},
\eeq
where $a$    is the  perturbation theory parameter
\beqn
a=
\frac{\alpha_s N_c}{2\pi}
\eeqn
and  $E_{\nu,n}$ is  the eigenvalue of the  BFKL Kernel in the adjoint representation   given by
\beqn\label{Enun}
E_{\nu,n}=-\frac{1}{2}\frac{|n|}{\nu^2+\frac{n^2}{4}}+\psi\left(1+i\nu+\frac{|n|}{2}\right)+\psi\left(1-i\nu+\frac{|n|}{2}\right)-2\psi(1).
\eeqn
Here $\psi(z)=\Gamma'(z)/\Gamma(z)$ and  $\gamma=-\psi(1)$ is the Euler constant.
The two loop LLA expression for remainder function in the BFKL approach was first found  from (\ref{corLLA}) and (\ref{LLA}) in ref.~\cite{BLS2}
\beqn\label{R2LLABFKL}
R^{(2)\; LLA}_{BFKL}\simeq\frac{i\pi}{2} \ln(1-u_1) \ln \tilde{u}_2\ln \tilde{u}_3= \frac{i\pi}{2} \ln(1-u_1) \ln\left|1+w\right|^2 \ln\left|1+\frac{1}{w}\right|^2.
\eeqn
This result was shown by Schabinger~\cite{Schabinger} to agree numerically with the expression obtained by analytic continuation of the remainder function found by Drummond, Henn, Korchemsky and Sokatchev~\cite{hexagon} from  Wilson Loop/Scattering Amplitude duality.
 The remainder function  (\ref{R2LLABFKL}) was then explicitly confirmed by two of the authors~\cite{LP1} performing the analytic continuation of the Goncharov-Spradlin-Vergu-Volovich~(GSVV)  two-loop expression~\cite{GSVV}.  The analytic continuation allowed also to extract the next-to-leading contribution, not yet available from the BFKL approach
\begin{eqnarray}\label{R2NLLABFKL}
 &&R^{(2)\;NLLA}_{}\simeq
 \frac{i\pi}{2} \ln |w|^2 \ln^2|1+w|^2-\frac{i\pi}{3}\ln^3 |1+w|^2+ i\pi \ln |w|^2 \left( \text{Li}_2 (-w) +\text{Li}_2 (-w^*)\right)
\nonumber \\
&&-i2\pi  \left( \text{Li}_3 (-w) +\text{Li}_3 (-w^*)\right).
\end{eqnarray}
The LLA term in (\ref{R2LLABFKL}) is pure imaginary and symmetric under  $w \to 1/w$ transformation in accordance with (\ref{LLA}). The next-to-leading~(NLLA) contribution in (\ref{R2NLLABFKL}) is also pure imaginary and has the same symmetry. Both of the contributions are pure imaginary due to a cancellation of the real part coming from the Mandelstam cut, Regge pole and a phase present in the BDS amplitude as was shown by one of the authors~\cite{LipDisp}. Starting at three loops this cancellation does not happen anymore and the real part gives a non-vanishing contribution at the next-to-leading level. The analysis of ref.~\cite{LipDisp} based on analyticity  and  other general properties of the scattering amplitudes resulted in a formulation of the dispersion-like relation for the real and imaginary parts of the remainder function in the Regge kinematics  at the Mandelstam region
\beqn\label{disp}
R \,e^{i\pi \,\delta}=\cos \pi \omega _{ab}+i\int _{-i\infty}^{i \infty}\frac{d\omega}{2\pi i}\,
f(\omega )\,e^{-i\pi \omega}\,(1-u_1)^{-\omega}\,,
\eeqn
where the first term in RHS corresponds to the contribution of the Regge pole.
This term  as well as the phase $\delta$ in LHS of (\ref{disp}) are obtained directly from the BDS formula~\cite{LipDisp}
\beqn\label{deltaomega}
\delta =\frac{\gamma _K}{8}\,\ln (\tilde{u}_2 \tilde{u}_3)\,,\,\,
\omega _{ab}
=\frac{\gamma _K}{8}\,\ln \frac{\tilde{u}_2}{\tilde{u}_3}.\,
\eeqn
The second terms in RHS of (\ref{disp})  stands for the contribution of the Mandelstam cut.
The coefficient  $\gamma _K \simeq 4 a$ is the cusp anomalous dimension known to an arbitrary  order of the perturbation theory. The only unknown piece in Eq.~\ref{disp} is the real function $f(\omega)$, which contains the Mandelstam cut in $\omega$ and depends only on the transverse particle momenta and has no energy dependence. In the leading logarithmic approximation $f(\omega)$ can be extracted from (\ref{LLA}) and reads
\beqn\label{fomegaLLA}
&&f^{LLA}(\omega)=\frac{a}{2}\sum_{n=-\infty}^{\infty} \int d \nu \frac{1}{\omega-\omega(\nu,n)} \frac{(-1)^n}{\nu^2+\frac{n^2}{4}}\left(w^* \right)^{i\nu -\frac{n}{2}}\,
\left(w\right)^{i\nu +\frac{n}{2}},
\eeqn
where $\omega(\nu,n)$ is defined in (\ref{eigen}).

The dispersion-like relation in (\ref{disp})  was used for calculating the three loop contributions to $R^{(3)}_6$~(leading imaginary and the sub-leading real terms) in the multi-Regge kinematics
\beqn\label{R3LLABFKL}
&&  R_{BFKL}^{(3)\;LLA}=i\Delta^{(3)\; LLA} _{2\rightarrow 4}/a^3=i\pi \frac{1}{4} \ln^2(1-u_1)\left(
 \ln|w|^2\ln^2|1+w|^2-\frac{2}{3}\ln^3|1+w|^2\right. \hspace{1cm}\;\;\;
\\
&&\left.-\frac{1}{4}\ln^2|w|^2 \ln|1+w|^2+\frac{1}{2} \ln|w|^2 \left(\text{Li}_2(-w)+\text{Li}_2(-w^*)\right)
 - \text{Li}_3(-w)-\text{Li}_3(-w^*)\right) \nonumber
\eeqn
and
\beqn\label{R3NLLABFKL}
&&  \Re\left(R^{(3)\;NLLA}_{BFKL}\right)= \frac{\pi^2}{4} \ln (1-u_1)\left(
 \ln|w|^2\ln^2|1+w|^2-\frac{2}{3}\ln^3|1+w|^2\right. \hspace{1cm}\;\;\;
\\
&&\left.-\frac{1}{2}\ln^2|w|^2 \ln|1+w|^2- \ln|w|^2 \left(\text{Li}_2(-w)+\text{Li}_2(-w^*)\right)
 +2 \text{Li}_3(-w)+2\text{Li}_3(-w^*)\right). \nonumber
 \eeqn
As in the two loop case,  both (\ref{R3LLABFKL})  and (\ref{R3NLLABFKL}) are symmetric under $w \to 1/w$ transformation, which is obvious from (\ref{LLA}) and corresponds to the target-projectile symmetry of the scattering amplitude. The  corrections, subleading in the logarithm of the energy,  are not captured by (\ref{LLA}) and require some knowledge of the next-to-leading impact factor and the intercept of the  BFKL eigenvalue in the adjoint representation. While the latter is still to be found from the next-to-leading BFKL equation, the correction to the impact factor was obtained in ref.~\cite{LP2} extracting it from (\ref{R2NLLABFKL}). This result showed an intriguing relation between next-to-leading corrections to the  impact factor at two loops and the three loop leading logarithmic contribution (see sections 4 and 5 of ref.~\cite{LP2} for more details).

In the next section we discuss the collinear limit of the scattering amplitudes in the multi-Regge kinematics, which is similar to  the double logarithmic approximation with  an overlapping of the BFKL and DGLAP   approaches.

\section{Collinear and Regge kinematics}\label{sec:collregge}
 In this section we consider the collinear limit of the amplitudes in the multi-Regge kinematics. In this limit two neighboring particles become collinear and one of the energy invariants tends to zero. Among a variety of possibilities we pick up one case, where the initial particle  with momentum $p_B$ in Fig.~\ref{fig:2to4ope} is collinear to a particle in the final state with momentum $p_{B'}$. This corresponds to $t_3\to 0$ and thus to $u_2 \to 0$. At this point we should take care about one fine point. Namely, in our analysis based on the BFKL approach we choose the largest scale dictated by the multi-Regge kinematics,  which produces the leading logarithms in each order of the perturbation theory in the effective summation parameter, which is $a \ln(1-u_1)$. Taking the collinear limit we introduce another, a potentially larger   parameter, which at the first sight does not satisfy the basic assumptions of the BFKL approach. In a general case the collinear and Regge limits do not necessarily commute, but having a physical intuition from the BFKL and DGLAP equations we have all reasons to believe that these two limits are interchangeable. This is indeed the case as will be shown later.

We start with taking $u_2 \to 0$ ($t_3\to 0$) faster than $1-u_1$, in other words we assume that the reduced cross ratio $\tilde{u}_2$ in (\ref{redcross}) vanishes  in contrast to the multi-Regge kinematics in (\ref{multicross}), where it is kept to be of the order of unity. It is also known that in the collinear kinematics $u_3  \simeq 1-u_1$ for $u_2 \to 0$ and we will use that fact later.  So that now we choose the following kinematics in terms of the dual conformal cross ratios~(compare to the Regge kinematics in (\ref{multicross}))
\beqn\label{collmulticross}
1-u_1 \to +0,\;\; u_2 \to +0, \;\; u_3 \to +0, \;\; \frac{u_2}{1-u_1}=\tilde{u}_2 \to +0, \;\; \frac{u_3}{1-u_1}=\tilde{u}_3 \simeq 1,
\eeqn
which in terms of $w$ and $w^*$ implies~(see (\ref{wdef}))
 \beqn\label{collw}
1-u_1 \to +0,\;\; |w|\to +0, \;\; \cos (\phi_2-\phi_3) \simeq \mathcal{O}(1).
\eeqn
In the BFKL approach  one sums large logarithms $\ln (1-u_1)$ keeping $|w|$ finite. We approach the  limit (\ref{collw})  by taking $\ln |w|$ to be of the order~(though not larger) of $\ln (1-u_1)$, which is still compatible with the BFKL resummation.   

 Expanding in this limit the two- and three-loop results for the remainder function in (\ref{R2LLABFKL}), (\ref{R2NLLABFKL}), (\ref{R3LLABFKL}) and (\ref{R3NLLABFKL}) we obtain
\beqn\label{coll2BFKL}
R^{(2)\;LLA}_{BFKL}+R^{(2)\;NLLA}\simeq -i2\pi \cos (\phi_2-\phi_3) \;|w|\;(\ln (1-u_1) \ln |w|+2 \ln |w|
-2)\;\;\;
\eeqn
and
\beqn\label{coll3LLABFKL}
R^{(3)\;LLA}_{BFKL}\simeq -\frac{i\pi}{2} \ln^2 (1-u_1) \cos (\phi_2-\phi_3) \;|w|\;\ln^2 |w|
\eeqn
as well as
\beqn\label{coll3RealBFKL}
\Re \left(R^{(3)\;NLLA}_{BFKL}\right)\simeq -\pi^2 \ln (1-u_1) \cos (\phi_2-\phi_3)\;|w|\; \ln^2 |w|.
\eeqn

The BFKL approach allows to calculate in the double logarithmic limit the leading imaginary and   real   contributions to any order of the perturbation theory.
 In the collinear limit $|w|\to 0$ the largest contribution in the integral over $\nu$ in (\ref{LLA}) comes from the poles at $\nu = -i n/2$ for $n=1$.
 The details of this calculation are presented in appendix~\ref{app:bessel} and the result is expressed in terms of the  modified Bessel functions $I_k(z)$ as follows.
The contribution leading in both $\ln(1-u_1)$ and $\ln|w|$ reads
\beqn\label{DLLA}
R^{DLLA}_{BFKL}\simeq   i2\pi a \cos (\phi_2-\phi_3) \;|w| \left(1-I_0\left(2 \sqrt{a \ln |w| \ln(1-u_1)}\right) \right),
\eeqn
while the real part of the  contribution suppressed in the logarithm of the energy $\ln(1-u_1)$~(NDLLA term) is given by
\beqn\label{ReNDLLA}
&&\Re\left(R^{NDLLA}_{BFKL}\right)\simeq 2\pi^2 a^{3/2} \cos (\phi_2-\phi_3)\; |w| \;\ln|w| \frac{I_1\left(2 \sqrt{a \ln |w| \ln(1-u_1)}\right)}{\ln(1-u_1)} \\
&& +  4 \pi^2 a \cos (\phi_2-\phi_3) \;|w|\; \ln|w| \left(1-I_0\left(2 \sqrt{a \ln |w| \ln(1-u_1)}\right) \right)
-2\pi^2 a^2 \cos(\phi_2-\phi_3) |w| \ln |w|. \nonumber
\eeqn

The collinear limit of the remainder function for the six-gluon planar MHV amplitude was  earlier considered  by Alday, Gaiotto, Maldacena, Sever and Vieira~(AGMSV) in ref.~\cite{OPEmalda}.  They suggested that introducing the following parametrization of the dual conformal cross ratios
\beqn\label{sigmatauphi}
u_2=\frac{1}{\cosh^2 \tau}, \;\; u_1=\frac{e^{\sigma}\sinh \tau \tanh \tau }{2(\cos \phi+\cosh \tau\cosh \sigma )}, \;\;
u_3=\frac{e^{-\sigma}\sinh \tau \tanh \tau }{2(\cos \phi+\cosh \tau\cosh \sigma )}
\eeqn
one can write the remainder function in the collinear limit $\tau\to \infty $ in a rather compact  way
\beqn\label{ROPE}
R^{(\ell)}_{OPE} \sim \cos \phi \; e^{-\tau} \frac{(-1)^{\ell-1}\tau^{\ell-1}}{(\ell-1)!} \int dp \;e^{ip\sigma}  c^{0}(p) \gamma^{\ell-1}_1(p),
\eeqn
where $\ell$ is a number of loops
and $\gamma_1(p)$ is the anomalous dimension of high spin operators considered in the operator product expansion of ref.~\cite{OPEmalda}~(see~also~a~paper~of~Basso~\cite{Basso:2010in})
\beqn\label{gamma}
\gamma_1(p)=\psi \left(\frac{3}{2}+\frac{ip}{2}\right)+\psi \left(\frac{3}{2}-\frac{ip}{2}\right)-2 \psi \left(1\right).
\eeqn

The function $c^{0}(p)$ can  found from one loop~(i.e. the  BDS expression) and reads
\beqn\label{c0raw}
c^{(0)}(p) \propto \frac{1}{1+p^2} \frac{\pi}{\cosh \frac{\pi p}{2}}.
\eeqn
Indeed,  for  the  $2 \to 4$ amplitude we can write the BDS at one loop  up to irrelevant terms that depend on $\mu^2$ and $\epsilon$ as
\beqn\label{I6F6}
&&I_6+F_6 \simeq \frac{1}{2} \ln  s \ln  s_2-\frac{1}{2} \ln  s\ln  t_1 -\frac{1}{2} \ln  s \ln  t_3-\frac{1}{2} \ln  s_1
   \ln  s_2-\frac{1}{2} \ln  s_1 \ln  t_1
\\
&&
 +\frac{1}{2} \ln  s_1 \ln  t_3-\frac{1}{2} \ln  s_2
   \ln  s_3
+\frac{1}{2} \ln  s_3\ln  t_1 -\frac{1}{2} \ln  s_3\ln  t_3+\frac{3 \pi ^2}{4} +R_1,\nonumber
\eeqn
where $R^{(1)}$ is a function of only anharmonic ratios $u_i$
\beqn\label{R1}
R^{(1)}=-\frac{1}{2} \sum_{i=1}^{3}\left(\frac{1}{2}\ln^2 u_i +\text{Li}_2(1-u_i)\right).
\eeqn
In the collinear limit $\tau \to \infty$ we obtain
\beqn
R^{(1)} \simeq -\tau ^2+2 \tau  \ln 2-\frac{\pi ^2}{6}-\ln ^2 2-\sigma ^2+\cos \phi \;e^{-\tau} h_0 (\sigma),
\eeqn
where
\beqn\label{h0}
h_0 (\sigma)= \int_{-\infty}^{\infty} c^{(0)}(p)  e^{ip\sigma} dp
\eeqn
for
\beqn\label{cobds}
c^{(0)}(p)= \frac{2}{1+p^2} \frac{1}{\cosh \frac{\pi p}{2}}.
\eeqn
This simple one-loop analysis allows us to fix the normalization in (\ref{c0raw}). However, taking into account some ambiguity in expressing the finite part of the BDS formula in terms of the anharmonic ratios we fix the normalization of $c^{(0)}(p)$ using the two-loop  remainder function.  Both of them give the same expression for $c^{(0)}(p)$ as we show later.  It is worth emphasizing that  the one-loop BDS ``remainder function``  $R^{(1)}$  in (\ref{R1}) can be compactly written as
\beqn\label{RR1}
R^{(1)}=\frac{1}{2} \sum_{i=1}^{3} \text{Li}_2\left( 1-\frac{1}{u_i}\right).
\eeqn
Polylogarithmic  functions of the same argument appear also in the two-loop  remainder function  suggesting an intimate relation between the  BDS amplitude and its  corrections.

The expression in (\ref{ROPE}) was also shown to agree numerically~\cite{OPEmalda} at two loops with the result of Goncharov, Spradlin, Vergu and
Volovich~(GSVV)~\cite{GSVV}. The GSVV remainder function was derived for quasi-multi Regge kinematics~\cite{DelDuca:2009au,DelDuca:2010zg}, but it was argued to be valid also in a general kinematics for positive values of the dual conformal cross ratios. The analytic continuation in one of the dual conformal cross ratios, namely the one given by (\ref{u1cont}), with subsequent multi-Regge limit of (\ref{multicross}) reproduces the LLA BFKL result~\cite{LP1} for the physical Mandelstam region. Naturally, an important  question to be asked is whether or not one can perform a similar analytic continuation of the  AGMSV expression in (\ref{ROPE}) to find an  agreement or disagreement  with the BFKL analysis. In the attempt of answering this question we immediately face a difficulty of treating the cosine factor in (\ref{ROPE}).
In deriving the remainder function in the collinear limit (\ref{ROPE}) it was assumed in ref.\cite{OPEmalda}  that the absolute value of $\cos \phi$ is finite and is much smaller than $\tau$. It is indeed the case also in the collinear and  Regge kinematics we are  interested in (see  (\ref{collmulticross})).
 However, it is easy to see from the definition
\beqn\label{cosPhi}
\cos \phi =\frac{u_1+u_2+u_3-1}{2 \sqrt{u_1 u_2 u_3}},
\eeqn
that in the course of the analytic continuation  (\ref{u1cont}) at $u_1=|u_1| e^{-i\pi}$  the numerator becomes of the order of $2$, while the denominator is still small and thus (\ref{cosPhi}) is not limited anymore.  This means that one cannot directly apply the analytic continuation (\ref{u1cont}) to the  AGMSV expression, because   the unlimited growth of $\cos \phi $ at $u_1=|u_1| e^{-i\pi}$ does not satisfy the assumptions of the collinear expansion and therefore (\ref{ROPE})
is not always valid  during  the analytic continuation.
 We face a similar problem performing the analytic continuation (\ref{u1cont}) of the GSVV remainder function, when the value of $1-u_1$ at  $u_1=|u_1| e^{-i\pi}$  becomes of the order of 2 and thus does not satisfy the first condition in the  multi-Regge kinematics given by (\ref{multicross}). However, the GSVV expression is valid for all positive values~(arbitrary kinematics) of the dual conformal cross ratios and therefore its continuation does not lead to any difficulty.

In the case of the collinear expansion the condition of having $\cos \phi$ being limited along the path of the analytic continuation forces us to modify the simple circular path for $u_1$ in (\ref{u1cont}) and/or change also paths of $u_2$ and $u_3$, which are trivial in (\ref{u1cont}). The initial and the final points of the analytic continuation should be the same, and a deformation of the continuation path is possible under condition that we do not cross any singularities of the remainder function.

It is plausible that one can   deform the path of the analytic continuation of the remainder function in  such a way  that $\cos \phi$ in (\ref{ROPE}) remains limited along a new path. In other words the new path could be compatible with the collinear kinematics.  To prove it in the case of the two-loop GSVV expression one should consider the analytic
continuation of the function of two variables $u_1$ and $u_3$, keeping $u_2$ fixed and small. We hope to do this in the future. Below we assume that such
a deformation of the path of the analytic continuation does exist.

Note that in general kinematics we defined~(see (\ref{sigmatauphi}))
 \beqn\label{u1u3sigma}
 \frac{u_1}{u_3}=e^{2\sigma}, \;\; \sigma =\frac{1}{2} \ln \frac{u_1}{u_3}
 \eeqn
 and thus the analytic continuation of (\ref{ROPE}) along  a path given by (\ref{u1cont}) in the complex  $\sigma$-plane would mean a simple shift
 \beqn\label{contA}
 \sigma \Rightarrow \sigma -i \pi,
 \eeqn
 where $\sigma$ is large and  positive for multi-Regge kinematics given in (\ref{multicross}).  We name this path in the $\sigma$-space \emph{the path $\mathbf{A}$} and argue that this continuation is not valid for expressions, where the collinear limit was performed first.

 On the other hand in the collinear kinematics (provided $\cos \phi$ is of the order of unity) the cross ratios can be approximated by~(see (\ref{sigmatauphi}))
\beqn
u_2 \simeq 4 e^{-2\tau}, \;\; u_1  \simeq \frac{e^{\sigma}}{2 \cosh \sigma}, \;\; u_3  \simeq \frac{e^{-\sigma}}{2 \cosh \sigma}
\eeqn
with a simple relation $u_3 \simeq 1-u_1$.  So that we can plug this relation in (\ref{u1u3sigma}) and redefine
\beqn\label{u1u1sigma}
 \sigma \simeq \frac{1}{2} \ln \frac{u_1}{1-u_1},
\eeqn
which gives the same expression for the function of  $\sigma$ in (\ref{ROPE}) in the  collinear  limit,
 but changes the path of the analytic continuation in the $\sigma$-plane for $u_1=|u_1| e^{-i2\pi}$. In fact, the use of (\ref{u1u1sigma}) instead of (\ref{u1u3sigma})
 in the expression (\ref{ROPE}) means the redefinition of this expression in the region beyond the collinear limit. We believe, that in the case, when $\sigma$ depends
 only on $u_1$, it is possible to prove, that the path of the analytic continuation can be deformed in such a way that $\cos \phi$ remains restricted along this path and, as a result,   (\ref{ROPE}) can be used for the continuation.

\begin{figure}[htbp]
	\begin{center}
		\epsfig{figure=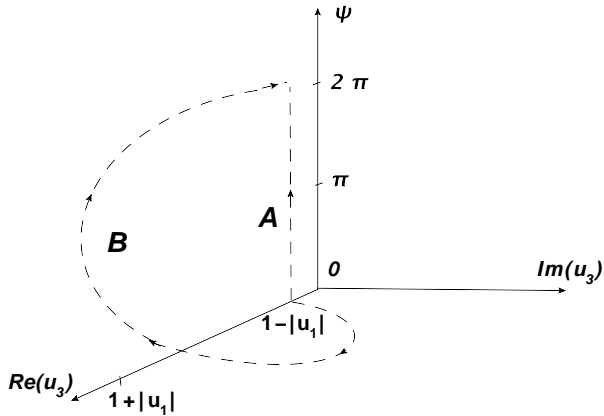,width=80mm}
	\end{center}
	\caption{ The paths $\mathbf{A}$ and $\mathbf{B}$ in the $u_3$-space. The phase $\psi$ is defined by $u_1=|u_1| e^{-i\psi}$ and takes values between $0$ and $2\pi$ in the course of the analytic continuation. The value of $u_3$ is the same  at the initial and final points of the paths $\mathbf{A}$ and $\mathbf{B}$.
 }
	\label{fig:spiral}
\end{figure}

 We name the  path corresponding to our new definition of $\sigma$ in (\ref{u1u1sigma}) \emph{the path $\mathbf{B}$}, and assume that the  analytic continuation with this deformed path is valid also for expression where the collinear limit was performed first. In particular, we see that in contrast to the analytic continuation along the path $\mathbf{A}$ the cosine factor can be made finite  at the point $u_1=|u_1|e^{-i\pi}$ since the numerator can have the same smallness as the denominator provided $u_3$ is adjusted in the corresponding way. Note that  this analytic continuation is different from one applied to the GSVV remainder function in ref.~\cite{LP1}, because $u_3$ is not kept fixed anymore and acquires some phase. However, in the course of  the analytic continuation along the path $\mathbf{B}$ we never cross the imaginary axis in the complex $u_3$-plane, i.e. never go to the negative real values of $u_3$ and thus do not cross the singularities of the GSVV expression.  The paths $\mathbf{A}$ and $\mathbf{B}$ in the $u_3$-space are illustrated in Fig.~\ref{fig:spiral}, where $\psi$ is defined by $u_1=|u_1| e^{-i\psi}$ and changes from $0$ to $2\pi$ in the course of the analytic continuation.

 The fact that $u_3$ never crosses the imaginary axis  allows us to deform smoothly the path of the continuation and so that it could be compatible with the collinear kinematics. To demonstrate this fact,  we  performed the analytic continuation (\ref{u1cont}) of the GSVV remainder function for arbitrary, but small values of $u_2$. Then  we took the collinear limit $u_2 \to 0$ of the continued function,  substituted $u_3 \simeq 1-u_1$ for an arbitrary value of $u_1$~(i.e. no Regge limit) reproducing the result of the analytic continuation of (\ref{ROPE}) along the path $\mathbf{B}$ for  arbitrary positive $\sigma$. This way we show that the analytic continuation of AGMSV expression in (\ref{ROPE}) along path $\mathbf{B}$ is justified.

As the next step in our analysis we want to apply the analytic continuation $\mathbf{B}$ to the AGMSV expression in (\ref{ROPE}) at higher loops  and  after taking the  $\sigma \to +\infty$ limit  compare the result  to the one obtained in the BFKL approach. This requires a knowledge of the overall constant, which is possible to fix at two loops expanding the GSVV remainder function at $\tau \to \infty$.  We find that it can be written as
\beqn\label{collGSVV}
R^{(2)}_{GSVV}\simeq \cos \phi \; e^{-\tau} \left(-\tau h_1(\sigma)+h^{sub}_1(\sigma)\right)+\mathcal{O}\left(e^{-2\tau}\right),
\eeqn
where $h_1(\sigma)$ is a function calculated in ref.~\cite{OPEmalda}
\beqn\label{h11}
h_1(\sigma)=-2\cosh \sigma \left(2 \ln \left( 1+e^{2\sigma}\right)\ln \left( 1+e^{-2\sigma}\right)-4\ln \left(2 \cosh \sigma\right)\right)
-8 \sigma \sinh \sigma.
\eeqn
 The sub-leading in $\tau$ contribution we extract from the GSVV expression
\beqn\label{h1sub}
&& h^{sub}_1(\sigma)=-\frac{2}{3} \pi ^2 \sigma  \cosh \sigma-4 \sigma ^2 \cosh \sigma-\frac{8}{3} \sigma ^3 \cosh \sigma+4 \sigma ^2 \cosh \sigma \ln 2
\\ \nonumber
&&-8 \cosh \sigma \ln(2 \cosh \sigma)+\frac{2}{3} \pi ^2 \cosh \sigma \ln(2 \cosh \sigma)+4 \sigma ^2 \cosh \sigma \ln(2 \cosh \sigma) \\ \nonumber
&&+8 \cosh \sigma  \ln(2 \cosh \sigma)\ln 2+4 \cosh \sigma \ln^2(2 \cosh \sigma)-4 \cosh \sigma  \ln^2(2 \cosh \sigma)\ln 2\\ \nonumber
&&-\frac{4}{3} \cosh \sigma \ln^3(2 \cosh \sigma)+4 \cosh \sigma \text{Li}_3 \left(-e^{-2\sigma}\right)-8 \sigma  \sinh \sigma-8 \sigma   \sinh \sigma \ln 2.
\eeqn
The functions $h_1(\sigma)$ and $h^{sub}_1(\sigma)$ are symmetric in $\sigma \to -\sigma$ and vanish at $\sigma \to \pm \infty$.
The explicit expression for $h_1(\sigma)$ allows to fix the overall coefficient  of the AGMSV remainder function in (\ref{ROPE}). The normalization fixed using two-loop GSVV expression coincides with the normalization we fixed using only BDS one-loop expression~(see~(\ref{cobds}) and the text whereafter).
Thus we can  write the AGMSV remainder function~(\ref{ROPE}) in the exponential form
\beqn\label{ROPEexp}
R_{OPE} \simeq a \cos \phi \; e^{-\tau} \int_{-\infty}^{\infty} c^{0}(p)\left( e^{-a\tau \gamma_1(p)}-1\right) e^{ip\sigma} d p\simeq a^2 R^{(2)}_{OPE}+a^3 R^{(3)}_{OPE}+...
\eeqn
with
\beqn\label{c0}
c^{(0)}(p)=\frac{2}{1+p^2} \frac{1}{\cosh \frac{\pi p}{2}}.
\eeqn
Introducing
\beqn\label{hk}
h_{k}(\sigma)= \int_{-\infty}^{\infty} c^{(0)}(p)\; \gamma^k_1 (p)\;e^{ip \sigma} dp
\eeqn
the AGMSV remainder function (\ref{ROPEexp}) can be written as
\beqn
R_{OPE} \simeq a \cos \phi \; e^{-\tau} \sum_{k=1}^{\infty} \frac{(-a \tau)^k }{k!} h_{k}(\sigma).
\eeqn
 The knowledge of $h_1^{sub}(\sigma)$ gives a possibility of calculating the  one-loop correction to the "coefficient function" $c^{0}(p)$.
To find the next-to-leading correction to the remainder function in the collinear limit one needs also corrections to the anomalous dimensions $\gamma_1(p)$ in~(\ref{gamma}).

\begin{figure}[htbp]
	\begin{center}
		\epsfig{figure=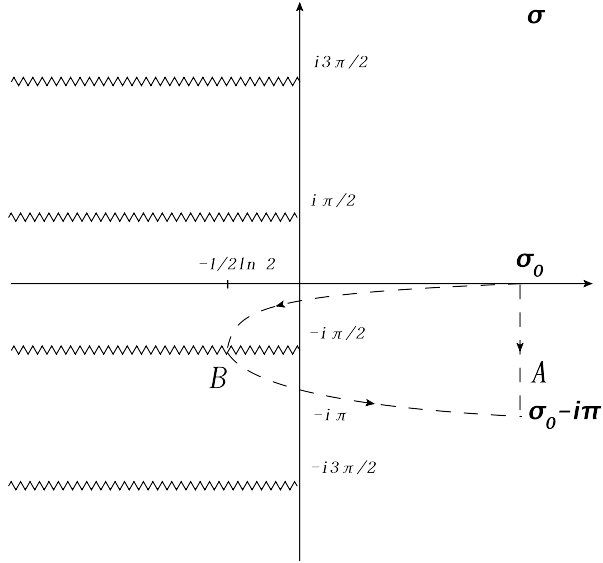,width=80mm}
	\end{center}
	\caption{ The cut structure of $h_k(\sigma)$ and $h^{sub}_1(\sigma)$. The figure illustrates the paths $\mathbf{A}$ and $\mathbf{B}$ of the analytic continuation.  $\sigma_0$ denotes some starting point of the analytic continuation. Both of the paths have the same starting and final  points. The Regge kinematics corresponds to $ \sigma_0 \to +\infty$.
 }
	\label{fig:cuts}
\end{figure}

Next we  investigate the analytic structure of the AGMSV remainder function in the complex $\sigma$-plane.
Both $h_1(\sigma)$ and $h^{sub}_1(\sigma)$ have the same branch cuts in the complex $\sigma$-plane starting at $\pm i \pi (2n+1)/2$ for $n=0,1,2,...$ as illustrated in Fig.~\ref{fig:cuts}.
From Fig.~\ref{fig:cuts} one can see the difference between the two paths $\mathbf{A}$ and $\mathbf{B}$ of the analytic continuation. The path $\mathbf{A}$ is  linear in the $\sigma$-plane and does not cross   horizontal branch cuts, while the path  $\mathbf{B}$ is non-trivial and does cross the  branch cut. For the purpose of the present discussion the main difference between the two cases is the fact that after the analytic continuation along the path $\mathbf{A}$ functions $h_k(\sigma)$ and $h_1^{sub}(\sigma)$ vanish for $\sigma \to +\infty$~(Regge limit), while for  the path $\mathbf{B}$ they give a non-vanishing term compatible with the logarithmic contributions in the BFKL approach.  For the details of the analytic continuation the reader is referred to  the appendix~\ref{app:cont} and here we present the main results.

We found that the analytic continuation along the path  $\mathbf{B}$ together with the  subsequent Regge limit $\sigma \to +\infty$ of the AGMSV remainder function (\ref{collGSVV}) fully reproduce the  BFKL result at two loops for the double (collinear and Regge ) logarithmic limit of the remainder function given by (\ref{coll2BFKL}). In particular, we establish a connection between the conformal spin defined in ref.~\cite{OPEmalda} and the conformal spin used in the BFKL approach noting that in the multi-Regge kinematics~(see (\ref{wdef}) and (\ref{cosPhi}))
\beqn
\cos \phi \simeq -\cos (\phi_2-\phi_3),
\eeqn
 which after the analytic continuation becomes
\beqn
\cos \phi \Rightarrow -\cos \phi \simeq \cos (\phi_2-\phi_3).
\eeqn

 To justify our  guess for the analytic continuation along the path $\mathbf{B}$ we perform the well-grounded analytic continuation (\ref{u1cont})
  for a general kinematics~(used in confirming the BFKL result in the multi-Regge kinematics) of the Goncharov-Spradlin-Vergu-Volovich~(GSVV) remainder function. Then we expand the continued GSVV expression for
 $\tau \to \infty$ and find that it coincides with  the result of the analytic continuation along the path $\mathbf{B}$ of the AGMSV remainder function
  for a positive fixed value of the parameter $\sigma$~(i.e. not necessarily in the Regge kinematics, where  $\sigma \to +\infty$). This confirms the validity of the continuation along the path $\mathbf{B}$ as well as it shows
 commutativity of the collinear and Regge limits.  One can expect the commutativity of these two limits from comparison of the BFKL~(Regge kinematics) and DGLAP~(Bjorken kinematics) equations, which   coincide in the double logarithmic limit.

 Using the definition of $h_k (\sigma)$ in (\ref{coll2BFKL}) we calculate analytically the AGMSV remainder function at three loops $R^{(3)}_{OPE}$ in (\ref{ROPEexp})~(see appendix~\ref{app:h2} for more details)
\beqn \label{h2}
&&h_2(\sigma)= \int_{-\infty}^{\infty}c^{0}(p)\gamma^2_1(p) e^{ip\sigma} dp=-\frac{\pi^2}{3} e^{-\sigma }-4 e^{-\sigma }   \sigma ^2-\frac{2}{3} \pi ^2 \sigma  \cosh \sigma
\\
&&+16  \sigma ^2 \cosh \sigma+\frac{8}{3}  \sigma ^3 \cosh \sigma+24  \cosh \sigma \ln (2 \cosh \sigma) \nonumber
+\frac{2}{3} \pi ^2 \cosh \sigma \ln (2 \cosh \sigma)
\\
&&
-8  \sigma ^2 \cosh \sigma \ln (2 \cosh \sigma)
-16  \cosh \sigma \ln^2 (2 \cosh \sigma) \nonumber
+\frac{16}{3}   \cosh \sigma \ln^3 (2 \cosh \sigma)
\\
&&
+8  \sigma  \cosh \sigma \;\text{Li}_2(-e^{-2\sigma})+8  \cosh \sigma \; \text{Li}_3(-e^{-2\sigma}) \nonumber
-24  \sigma  \sinh \sigma
+4  \sinh \sigma \; \text{Li}_2(-e^{-2\sigma}).  \nonumber
\eeqn

The analytic continuation along the path $\mathbf{B}$ of $R^{(3)}_{OPE}$  reproduces the BFKL remainder function in the double logarithmic approximation at three loops given by (\ref{coll3LLABFKL}) and  (\ref{coll3RealBFKL}), namely
\beqn\label{R3opecont}
&& R^{(3)}_{OPE} \overbrace{\Longrightarrow}^{\text{path} \mathbf{B}}-\frac{i\pi}{2} \ln^2(1-u_1)\cos (\phi_2-\phi_3)\;|w|\; \ln^2|w|-\pi^2 \ln(1-u_1)\cos (\phi_2-\phi_3)\;|w|\; \ln^2|w|\nonumber\\
&&
\hspace{2cm}-i3\pi \ln(1-u_1)\cos (\phi_2-\phi_3)\;|w|\; \ln^2|w|.
\eeqn

The first two terms in RHS of (\ref{R3opecont}) coincide with the corresponding BFKL expressions in (\ref{coll3LLABFKL}) and  (\ref{coll3RealBFKL}), while the last term is currently not accessible in the BFKL analysis and brings some new information about the next-to-leading eigenvalue of the BFKL eigenvalue in the adjoint representation.  For higher loops we need an analytic form of $h_k(\sigma)$, which are not considered here due to the complexity of the calculations for $k>2$.

The analytic continuation along the path $\mathbf{B}$  for the AGMSV remainder function is technically more involved than a simple shift $\sigma \Rightarrow \sigma -i\pi$ in the path  $\mathbf{A}$. Despite the fact that the continuation along the path $\mathbf{A}$ is not applicable for the AGMSV remainder function  as we discussed
earlier, we still can make a use of it at higher loops due to an interesting transformation property of $h_k(\sigma)$
\beqn\label{contAB}
\begin{tabular}{ l  l }
  \text{\emph{continuation along the path } $\mathbf{A}$}: & $h_k(\sigma) \Rightarrow -h_k(\sigma)+\Delta_{k}^* (-\sigma)$ \\
  \text{\emph{continuation along the path } $\mathbf{B}$}:  &  $h_k(\sigma) \Rightarrow -h_k(\sigma)+\Delta_{k} (\sigma)$,  \\
\end{tabular}
\eeqn
 where  $\Delta_{k} (\sigma)$ is some function of $\sigma$ calculated in the appendix~\ref{app:cont} for $k=1$ and $k=2$, which corresponds to two and three loops of the AGMSV remainder function. We checked that analytic
 continuation along  the path $\mathbf{B}$ of $h_k(\sigma)$ can be obtained by complex conjugation and reversing the argument of the analytic continuation along the  path $\mathbf{A}$
  and vice versa at two and three loops. We believe that this property holds at higher loops as well.
It is technically much easier to perform the analytic continuation with the path $\mathbf{A}$ and then calculate from it the required function $\Delta_{k} (\sigma)$ using the property (\ref{contAB}).

The minus sign of $h_k(\sigma)$ that appears on RHS of (\ref{contAB}) is related to the fact that after the analytic continuation $\cos \phi$ also changes the sign
\beqn
\cos \phi \Rightarrow -\cos \phi
\eeqn
so that the AGMSV remainder function (\ref{ROPEexp}), which is a product of $h_{k}(\sigma)$ and $\cos\phi$, does not  change the sign and only gets an additive discontinuity  $-\cos \phi \;\Delta_{k} (\sigma)$ as expected.

 We also made another intriguing observation, namely that
   the function $\Delta_{k} (\sigma)$ is much simpler than  $h_k(\sigma)$ and can be obtained from
the expression~(\ref{hk}) with omitted $\cosh (\pi p/2)$ in the denominator
   \beqn\label{Fk}
   F_k(\sigma)=\int_{-\infty}^{\infty} c^{(0)}(p) \;\gamma_1^{k}(p) \left(2 \cosh\frac{\pi p}{2}\right) \;e^{ip\sigma} dp =
   \int_{-\infty}^{\infty} \; \frac{4}{1+p^2} \gamma_1^{k}(p) \;e^{ip\sigma} dp
   \eeqn
     by  changing the sign of $\sigma$ and shifting it  by $i\pi/2$
     \beqn \label{DeltaF}
     \Delta_{k}(\sigma)=F_k \left(-\sigma+\frac{i \pi}{2}\right).
     \eeqn
We have checked this property at two and three loops, i.e. for
\beqn
&& F_{1}(\sigma)=8 e^{-\sigma } \pi +8 \pi  \ln(1-e^{-2\sigma})\sinh \sigma, \\
&& F_{2}(\sigma)=24 e^{-\sigma } \pi -8 e^{-\sigma } \pi  \sigma +32 \pi  \ln\left(1-e^{-2 \sigma}\right) \sinh \sigma-16 \pi  \sigma  \ln\left(1-e^{-2 \sigma}\right) \sinh \sigma \nonumber\\
&&\hspace{1cm}-16 \pi  \ln^2\left(1-e^{-2 \sigma}\right) \sinh \sigma-8 \pi \; \text{Li}_2 \left(-e^{-2 \sigma}\right) \sinh \sigma . \nonumber
\eeqn

 The main advantage of this observation is that $F_k(\sigma)$  is much easier  to calculate than the initial function $h_{k}(\sigma)$.
 We believe that the two properties (\ref{contAB}) and (\ref{DeltaF}) are intimately related to each other and their possible interpretation in terms of the energy discontinuities is presented in section~\ref{sec:reggeinter}.

\subsection{Four and five loops}\label{sec:45text}
In the previous part of the paper we performed the  analytic continuation of the AGMSV remainder function (\ref{ROPEexp}) at two and three loops, and then taking the Regge limit we reproduced the BFKL result (\ref{DLLA}) and (\ref{ReNDLLA}) in the Double Leading Logarithmic Approximation~(DLLA).  In doing this we needed an explicit analytic form of the function $h_{k} (\sigma)$. Going beyond three loops (i.e. $h_2(\sigma)$) in (\ref{ROPEexp}) presents a technical challenge and we found it much easier to calculate contributions only of powers of  $\gamma_{1}^{\pm}(p)$ defined by
\beqn\label{gammaminus}
\gamma_1(p)=\gamma^{+}_1(p)+\gamma^{-}_1(p), \;\;\;\gamma^{\pm}_1(p)=\psi \left(\frac{3}{2}\pm\frac{ip}{2} \right)-\psi \left(1 \right).
\eeqn
A possible physical interpretation of  the functions  $\gamma^{\pm}_1(p)$ is discussed in the next section.
 It turns out that the main contribution to the AGMSV remainder function in DLLA in the Mandelstam channel comes from the maximal powers of $\gamma_{1}^{-}(p)$ in
(\ref{ROPEexp}). Each power of $\gamma_{1}^{+}(p)$ introduced a suppression by one power of $\ln (1-u_1)$ in terms, which are  leading in $\ln |w|$.   Generally, we consider the multi-loop contribution from $\gamma_{1}^{+}(p)$  and $\gamma_{1}^{-}(p)$ in separate, introducing
\beqn\label{hkPM}
h^{+,...,-}_k (\sigma)=\int_{-\infty}^{\infty} c^{0}(p) \left( \gamma^{+}_1(p) \right)^{m} \left( \gamma^{-}_1(p) \right)^{k-m} e^{ip\sigma} dp,
\eeqn
where $m$ is the number of powers of  $\gamma^{+}_1(p)$ and $k=\ell -1$ is related to a number of loops $\ell$.
 The functions  $h^{\pm}_k (\sigma)$ are calculated in   appendix~\ref{app:45plus} and   given  for the two-loop case by
\beqn\label{h1minus}
&&
h^{-}_1 (\sigma)=\int_{-\infty}^{\infty} c^{0}(p)  \gamma^{-}_1(p) e^{ip\sigma} dp=
4 e^{-\sigma } \sigma -\frac{1}{3} \pi ^2 e^{-\sigma }
+4  \cosh \sigma \ln \left(1+e^{-2 \sigma }\right)
 \\
&&
 -2 \cosh \sigma \ln ^2\left(1+e^{-2 \sigma }\right)
   -4 \cosh \sigma \text{Li}_2\left(-e^{-2 \sigma }\right) \nonumber
\eeqn
and
\beqn\label{h1kP}
&& h^{+}_1 (\sigma)=h_1 (\sigma)-h^{-}_1 (\sigma)=4 \cosh  \sigma \text{Li}_2\left(-e^{-2 \sigma }\right)  +6 \sigma ^2 \cosh  \sigma +4 e^{-\sigma } \sigma +\frac{1}{3} \pi ^2 e^{-\sigma }
\\
&&-4
   \sigma  \cosh  \sigma -2 \cosh  \sigma  \ln  ^2(2 \cosh  \sigma )-4 \sigma  \cosh  \sigma  \ln  (2 \cosh  \sigma )+4 \cosh  \sigma
   \ln  (2 \cosh  \sigma ). \nonumber
\eeqn
Note that it follows from the definitions (\ref{gammaminus}) and (\ref{hkPM}) that  they are related by $h^{+}_1 (\sigma)=h^{-}_1 (-\sigma)$.
Similarly to  the second line of (\ref{contAB}), these functions after analytic continuation along path $\mathbf{B}$  also  can be written as
 \beqn\label{DeltaPMdef}
h^{\pm}_k (\sigma) \Rightarrow -h^{\pm}_k (\sigma)+\Delta_k^{\pm}(\sigma),
\eeqn
where $\Delta_k^{\pm}(\sigma)$ read
\beqn\label{Delta1PM}
\Delta_{1}^{+} (\sigma)=-4 i \pi  e^{\sigma }, \;\;\Delta_1^{-}(\sigma) =-4 i \pi  e^{\sigma }+8 i \pi  \sigma  \cosh \sigma +8 i \pi  \cosh \sigma  \ln (2 \cosh \sigma ) .
\eeqn
In the multi-Regge kinematics $\sigma \to \infty$ we get their respective contributions to the remainder function
\beqn\label{R2P}
&&R^{(2)+}_{OPE} =-\cos \phi e^{-\tau} \tau h^{+}_{1} (\sigma) \Longrightarrow
-i 2 \pi  \cos  (\phi_2-\phi_3)   |w| \ln  |w|
  \eeqn
and
\beqn\label{R2M}
 &&R^{(2)-}_{OPE} =   -\cos \phi \; e^{-\tau}\;\tau h^{-}_1 (\sigma) \Longrightarrow -i2  \pi  \cos  (\phi_2-\phi_3)  |w| \ln |w| \ln (1-u_1)
\\
&&
    \;\;\;\;\;\;\;\;-i2  \pi \cos  (\phi_2-\phi_3)   |w|\ln |w|. \nonumber
\eeqn
In (\ref{R2P}) and (\ref{R2M}) we omit  terms not enhanced by $\ln |w|$, which are irrelevant for the present discussion. The full form of  $R^{(2)\pm}_{OPE}$ is presented in appendix~\ref{app:45plus}. From  (\ref{R2P}) and (\ref{R2M}) we see that $R^{(2)+}_{OPE} $ is suppressed by one power of $\ln (1-u_1)$ with respect to $R^{(2)-}_{OPE} $.  The $R^{(2)+}_{OPE} $ contribution is subleading and  not captured by the double-logarithmic BFKL analysis. In order to find a corresponding contribution in the BFKL approach one needs to calculate the next-to-leading eigenvalue of the  BFKL  Kernel in the adjoint representation,  and it is not currently available.

At the three loop level we also observe a similar situation, where the remainder function (\ref{ROPEexp}) is given by
\beqn
R_{OPE}^{(3)}=R_{OPE}^{(3)++}+R_{OPE}^{(3)+-}+R_{OPE}^{(3)--}
\eeqn
with
\beqn
R_{OPE}^{(3)++}=\cos \phi \;e^{-\tau} \frac{\tau^2}{2} h^{++}_2 (\sigma) , \;\; R_{OPE}^{(3)+-}=\cos \phi \; e^{-\tau}  \tau^2  h^{+-}_2 (\sigma) ,\;\;
 R_{OPE}^{(3)--}=\cos \phi \;e^{-\tau} \frac{\tau^2}{2} h^{--}_2 (\sigma) .
\eeqn
In the Double Leading Logarithmic Approximation for the Mandelstam channel we obtain
\beqn\label{R3PPtext}
&& R_{OPE}^{(3)++} \Longrightarrow -i\pi   |w| \ln ^2|w|
\eeqn
and
\beqn\label{R3PMtext}
&&R_{OPE}^{(3)+-} \Longrightarrow -i2  \pi \cos  (\phi_2-\phi_3)   |w| \ln  (1-u_1) \ln  ^2 |w|+\frac{i \pi ^3}{3}   \cos  (\phi_2-\phi_3) |w| \ln  ^2|w|
\\
 &&-i4  \pi \cos  (\phi_2-\phi_3)  |w| \ln  ^2|w| \nonumber
\eeqn
as well as
\beqn\label{R3MMtext}
 &&R^{(3)--}_{OPE} \Longrightarrow
-\frac{i \pi}{2}  \cos  (\phi_2-\phi_3)   |w| \ln ^2|w| \ln ^2(1-u_1)
\\
&&-\pi ^2  \cos  (\phi_2-\phi_3)  |w| \ln ^2|w| \ln (1-u_1).   \nonumber
\eeqn
It is clear from (\ref{R3PPtext}), (\ref{R3PMtext}) and (\ref{R3MMtext}) that each power of $\gamma_1^{+}(p)$ brings an additional  suppression by one power of $\ln (1-u_1)$.  We expect this to happen also at higher loops and argue that the main contribution in DLLA in the Mandelstam channel comes from the maximal power of $\gamma_1^{-}(p)$ in (\ref{hkPM}), namely for $m=0$
\beqn\label{hkPPPPP}
h^{-,...,-}_k (\sigma)=\int_{-\infty}^{\infty} c^{0}(p) \left( \gamma^{-}_1(p) \right)^{k} e^{ip\sigma} dp,
\eeqn
which we call $h^{-}_k (\sigma)$ for short. The functions $h^{-}_3 (\sigma) $ and $h^{-}_4 (\sigma) $ , which corresponds to $4$ and $5$ loops respectively, were
calculated in appendix~\ref{app:45plus}.  The function $h^{-}_3 (\sigma) $ is given by
\beqn\label{h3minus}
&&
h^{-}_3 (\sigma)=\int_{-\infty}^{\infty} c^{0}(p) \left( \gamma^{-}_1(p) \right)^3 e^{ip\sigma} dp=
-\pi^2e^{-\sigma } +4 \sigma e^{-\sigma }+\frac{4}{15} \pi ^4 \cosh \sigma-e^{\sigma } \pi ^2 \ln \left(1+e^{-2\sigma}\right)
\\
&&+4 \cosh \sigma \ln \left(1+e^{-2\sigma}\right)
 -6 e^{-\sigma } \sigma  \ln^2 \left(1+e^{-2\sigma}\right)
-6 \cosh \sigma \ln^2 \left(1+e^{-2\sigma}\right)+12 \sigma  \cosh \sigma \ln^2 \left(1+e^{-2\sigma}\right) \nonumber
\\
&&+2 e^{-\sigma } \ln^3 \left(1+e^{-2\sigma}\right)+3 e^{\sigma } \ln^3 \left(1+e^{-2\sigma}\right) \nonumber
+4 \sigma  \cosh \sigma \ln^3 \left(1+e^{-2\sigma}\right)
-6 e^{-\sigma } \text{Li}_2\left(-e^{-2 \sigma}\right)
\\
&&-2 e^{\sigma } \text{Li}_2\left(-e^{-2 \sigma}\right)+6 e^{-\sigma } \ln \left(1+e^{-2\sigma}\right) \text{Li}_2\left(-e^{-2 \sigma}\right) \nonumber
-6 \cosh \sigma \ln^2 \left(1+e^{-2\sigma}\right) \text{Li}_2\left(-e^{-2 \sigma}\right)
\\
&&-4 \cosh \sigma \text{Li}_2\left(-e^{-2 \sigma}\right)^2-6 e^{-\sigma } \text{Li}_3\left(-e^{-2 \sigma}\right)-2 e^{\sigma } \text{Li}_3\left(-e^{-2 \sigma}\right)\nonumber
+4 \cosh \sigma \ln \left(1+e^{-2\sigma}\right) \text{Li}_3\left(-e^{-2 \sigma}\right)
\\
&&-6 e^{\sigma } \text{Li}_3\left(\frac{1}{1+e^{-2 \sigma}}\right)-12 \cosh \sigma \ln \left(1+e^{-2\sigma}\right) \text{Li}_3\left(\frac{1}{1+e^{-2 \sigma}}\right) \nonumber
-4 \cosh \sigma \text{Li}_4\left(-e^{-2 \sigma}\right)
\\
&&-24 \cosh \sigma \text{Li}_4\left(\frac{1}{1+e^{-2 \sigma}}\right)+12 \cosh \sigma \text{Li}_{2,2}\left(-e^{-2 \sigma}\right)+6 e^{\sigma } \zeta_3 \nonumber
-12 \cosh \sigma \ln \left(1+e^{-2\sigma}\right) \zeta_3, \nonumber
\eeqn
 while the expression  for $h^{-}_4 (\sigma) $ is rather lengthy and is presented in appendix~\ref{app:45plus}~(see (\ref{h4minusapp})) in terms of the harmonic polylogarithms~(HPL)~\cite{Moch:2001zr}.

The analytic continuation of  $h^{-}_3 (\sigma) $ and $h^{-}_4 (\sigma) $ along path $\mathbf{B}$ with the subsequent Regge limit allows us to find the OPE remainder function~(\ref{ROPEexp}) in the Mandelstam region in the Double Leading Logarithmic Approximation~(DLLA). At four and five loops they read
\beqn\label{R4M}
&& R^{(4)-}_{OPE} =   -\cos \phi \; e^{-\tau}\;\frac{\tau^3}{3!} h^{-}_3 (\sigma) \Longrightarrow
  -\frac{i \pi}{18} \cos (\phi_2-\phi_3)   |w| \ln ^3|w| \ln ^3(1-u_1)  \;\;\;\;\;\;\;\;\;\;
\\
&&-\frac{\pi ^2}{3} \cos (\phi_2-\phi_3)  |w|\ln ^3 |w| \ln ^2(1-u_1) -\frac{i\pi}{6} \cos (\phi_2-\phi_3)  |w| \ln ^3 |w|\ln
   ^2(1-u_1)  \nonumber
\eeqn
and
\beqn\label{R5M}
&& R^{(5)-}_{OPE} =   \cos \phi \; e^{-\tau}\;\frac{\tau^4}{4!} h^{-}_4 (\sigma) \Longrightarrow
-\frac{i \pi}{288} \cos  (\phi_2-\phi_3)    |w| \ln^4 |w| \ln^4(1-u_1)
\\
&& -\frac{ \pi ^2}{24}\cos  (\phi_2-\phi_3)  |w|  \ln^4 |w| \ln^3(1-u_1)
-\frac{i\pi}{72}  \cos  (\phi_2-\phi_3)  |w| \ln^4 |w| \ln^3(1-u_1).
\nonumber
\eeqn
The first two terms in RHS of the  remainder functions $R^{(4)-}_{OPE}$ and $R^{(5)-}_{OPE}$ reproduce the BFKL result in (\ref{DLLA}) and (\ref{ReNDLLA}), while the last term is beyond the applicability of the double-logarithmic BFKL analysis and requires a knowledge of the NLO impact factor, calculated by of the authors in ref.~\cite{LP2} as well as  the corrections to  the  eigenvalue of the BFKL Kernel   in the adjoint representation. This can be obtained from the NLO BFKL Kernel in the adjoint representation  found by Fadin and Fiore~\cite{Fadin:2004zq,Fadin:2005zj}.

 In this section we showed  that in order to reproduce known BFKL results in the double-logarithmic approximation up to five-loop level, it is enough to consider only a part of the anomalous dimension in (\ref{ROPEexp}). Namely, all of the leading terms come from $\gamma_{1}^{-}(p)$ in (\ref{gammaminus}), while each power of $\gamma_{1}^{+}(p)$ introduces a suppression in one power of $\ln(1-u_1)$ in the Mandelstam region.  In the next section we discuss this observation and argue that it could be a sign for a non-multiplicative renormalization of the remainder function in the collinear limit.

\section{Interpretation of the collinear limit from the Regge Theory }\label{sec:reggeinter}

The OPE expansion (\ref{ROPEexp}) for the remainder function  has a form of the Fourier integral transform in the variable $p$.  With the definition (\ref{u1u3sigma}) for $\sigma$ it is symmetric to the substitution $\sigma \rightarrow -\sigma$, which corresponds to the symmetry of the amplitude to the interchange of the cross ratios $u_1 \leftrightarrow u_3$. It is related to the symmetry of the Fourier transformed expression to the substitution $p \rightarrow -p $. Moreover, the function (\ref{ROPEexp})  can be analytically continued from the channel with $\sigma >0$ to the channel with $\sigma <0$ along the real axes, where it does not have any singularity. Note, that the channels with $\sigma>0$ and $\sigma<0$ are analogous to the $s$ and $u$-channels for the elastic~(nonplanar) amplitude.

However,  in the attempt to continue (\ref{ROPEexp}) to the Mandelstam region with $u_1 \simeq |u_1| e^{-i2\pi}$, one  faces some difficulties  as it was discussed in the previous section. To overcome them we suggested  to use another definition for $\sigma$~(see (\ref{u1u1sigma}))
\beqn\label{u1u1sigma2}
\sigma \simeq \frac{1}{2}\ln\frac{u_1}{1-u_1} , 
\eeqn
because in this case we could stay in the collinear limit with a fixed value of $\cos \phi$ in the course of the analytic continuation.  Note, that  the definitions (\ref{u1u3sigma}) and (\ref{u1u1sigma2}) are equivalent in the Euclidean collinear region, where $u_3 \simeq 1-u_1$, but the use of  (\ref{u1u1sigma2}) extends the region of applicability of the AGMSV  remainder function $R_{OPE}$ in (\ref{ROPEexp}). Note, that for the analytic continuation to the  Mandelstam region with $u_3 \to 1$ one should use the symmetric definition $\sigma=1/2 \ln ( (1-u_3)/u_3)$.

It turns out, that we have an analogous situation with the variable
\beqn
\tau \simeq \frac{1}{2} \ln \frac{4}{u_2},
\eeqn
which tends to infinity in the collinear limit. Indeed, in the Regge kinematics $\sigma \simeq 1/2 \ln s_2 \to \infty$, where according to the definitions (\ref{redcross})
\beqn
\tau \simeq \frac{1}{2} \ln \frac{4}{\tilde{u}_2} +\frac{1}{2} \ln \frac{1}{1-u_1} =\frac{1}{2} \ln \frac{4}{\tilde{u}_2} +\frac{1}{2} \ln \left(1+e^{2\sigma} \right)
\simeq \frac{1}{2} \ln \frac{4}{\tilde{u}_2} +\sigma
\eeqn
the variable $\tau$ depends on $\sigma$ for fixed $\tilde{u}_2$, the expression for $R_{OPE}$ in (\ref{ROPEexp}) would contain apart from the  large terms of the order of  $a \ln |w| \ln (1-u_1)$ also the comparatively large contributions $a \ln^2 (1-u_1)$, which are not in an agreement with the double-logarithmic asymptotics (\ref{DLLA}) obtained from the BFKL resummation. In principle these  contributions  can  be canceled by   higher-loop corrections to the  ''coefficient function'' $c^{0}(p)$. Indeed,  at  two loops the expression (\ref{h1sub}) for $h_1^{sub}(\sigma)$ contains the term $~\sigma^2$  which cancels exactly a similar term in the leading contribution $~\tau h_1(\sigma)$ in (\ref{collGSVV}). In our opinion such miraculous cancelations can be avoided, if one would redefine $\tau$ appearing in the powers $\ell-1$ in expressions of the type (\ref{ROPE})  in the following way
\beqn
\tau \rightarrow \frac{1}{2} \ln \frac{u_3}{u_2}=-\ln |w|,
\eeqn
where $|w|$ is given in (\ref{wdef}). In the collinear region this substitution  can be justified in the LLA $a \tau \sim 1, \; a \sigma \ll 1$, where the corresponding OPE formula (\ref{ROPE}) was derived. Note, that  the variables $|w|^{-2}$ and $1-u_1$  are analogous to the standard  Bjorken variables $\mathbf{Q}^2$ and $x$ in DIS. Thus, we suggest to write expression (\ref{ROPEexp})  in the collinear region matching the double-logarithmic limit as follows
\beqn\label{ROPEw}
R_{OPE} \simeq -a \cos (\phi_2-\phi_3) \frac{|w|}{2} e^{-\sigma}\int^{+\infty}_{-\infty} dp \;e^{i p\sigma} c^{0}(p) (|w|^{a \gamma_1(p)}-1),
\eeqn
where  $\cos (\phi_2-\phi_3) $ is defined in (\ref{wdef}) and differs from $\cos \phi$ given by  (\ref{cosPhi}) only by the factor $\sqrt{u_1}$ and the overall sign.

Now we  introduce the new variables
\beqn\label{pomega}
\sigma =\frac{1}{2} \ln \tilde{s}_2,\;  \tilde{s}_2=\frac{u_1}{1-u_1},\;  \;p=-i 2 \omega -i,
\eeqn
where $\tilde{s}_2$ is proportional to the invariant $s_2$ in Fig.~\ref{fig:2to4ope}. Note, that in the variable $\sigma$ the remainder function at one or two loops contained an essential singularity at infinity.  In the variable $\tilde{s}_2$ the essential singularity is absent.

\begin{figure}[htbp]
	\begin{center}
		\epsfig{figure=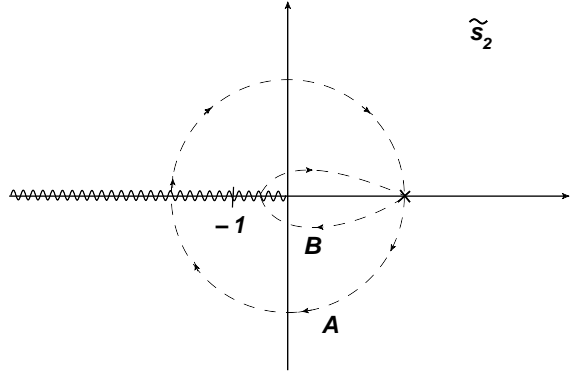,width=75mm}
	\end{center}
	\caption{ The paths $\mathbf{A}$ and $\mathbf{B}$ of the analytic continuation in the $\tilde{s}_2$-plane. The cross on the real axis denotes the initial  point of the continuation. In the course of the analytic continuation along the path $\mathbf{B}$ we cross the branch cut on the real axis from $0$ to $-1$ and return to the initial point. For the path $\mathbf{A}$ we cross   the cut from $-1$ to $-\infty$. }
	\label{fig:s2ABcuts}
\end{figure}

Using (\ref{pomega}) one can recast (\ref{ROPEw}) to  the Regge-like form

\beqn\label{ROPEomega}
R_{OPE } \simeq \frac{a \pi}{4} \cos( \phi_2-\phi_3) |w| \int_{-\frac{1}{2}-i\infty}^{-\frac{1}{2}+i\infty} \frac{ d\omega }{2 \pi i}
\frac{(\tilde{s}_2)^{\omega}}{\omega (\omega+1)}\frac{1}{\sin \pi \omega} \left(|w|^{a \tilde{\gamma}_{1}(\omega)}-1 \right),
\eeqn
where $\omega=j-1$   and
\beqn
\tilde{\gamma}_{1}(\omega)= \gamma_{1}(-i2\omega-i)=\psi (2+\omega)+\psi(1-\omega)-2\psi(1).
\eeqn
 The function $e^{\sigma}R_{OPE}$ is symmetric to the substitution $\tilde{s}_2\rightarrow 1/\tilde{s}_2$ and has singularities at the points~(cf.~Fig.~\ref{fig:cuts})
\beqn
\ln \tilde{s}_2 =\pm i\pi n
\eeqn
corresponding to the value $\tilde{s}_2=(-1)^n$, where the integral in (\ref{ROPEomega})   is divergent at large $\omega$.
Due to the symmetry of $e^{\sigma}R_{OPE}$ to the substitution $\sigma \rightarrow - \sigma$  the points $\tilde{s}_2=0$ and $\tilde{s}_2=-\infty$ are also singular and we can draw the cut in the $\tilde{s}_2$-plane  from $0$ to $-\infty$. The discontinuity of $R_{OPE}$ on this cut has a singularity at the point $\tilde{s}_2=-1$. Using formally its analytic continuation corresponding to the path $\mathbf{A}$ in Fig.~\ref{fig:cuts} for large positive $\sigma_0$,  we move along the large circle in a clockwise direction~(see~Fig.~\ref{fig:s2ABcuts} ) in the $\tilde{s}_2$-plane  and after crossing the cut at $\tilde{s}_2<-1$ return to the initial point. The difference between the values of $R_{OPE}$ after and before continuation in an accordance with the first equation of (\ref{contAB}) is equal to the discontinuity on this cut, analytically continued
from negative  to  the large positive $\tilde{s}_2$. We write the discontinuity before this continuation
\beqn\label{DeltaAs2}
\Delta R^{A}_{OPE} (-|\tilde{s}_2|)\simeq \frac{a \pi }{4} \cos (\phi_2 -\phi_3) |w| \theta(-1-\tilde{s}_2) \int_{-\frac{1}{2}-i \infty}^{-\frac{1}{2}+i \infty}\frac{d \omega}{2 \pi i}
 \frac{(-2i)\; \left|\tilde{s}_2\right|^{\omega} }{\omega (\omega+1)}  \left(|w|^{a \tilde{\gamma}_1(\omega)}-1\right) \;\;\;
\eeqn
Note, that the discontinuity on the cut at $\tilde{s}_2<-1$ is defined by a convergent integral, but its continuation $|\tilde{s}_2|\rightarrow e^{-i\pi} |\tilde{s}_2|$ to positive values of $\tilde{s}_2$ should be performed in a cautious way, because it demands the simultaneous rotation of the contour of integration by the angle $\pi$ in anti-clockwise direction, which is a rather complicated procedure  due to the infinite number of poles of the integrand.
On the other hand, using the correct analytic continuation of $R_{OPE}$, corresponding to the path $\mathbf{B}$ in the $\sigma$-plane of Fig.~\ref{fig:cuts}, we initially cross the cut at $-1<\tilde{s}_2<0$ moving from below and after that  return to  the initial point as illustrated in Fig.~\ref{fig:s2ABcuts}.
In this case the difference between  the values of $R_{OPE}$ after  and before  the continuation will be
\beqn\label{DeltaBs2}
\Delta R^{B}_{OPE} (-|\tilde{s}_2|)\simeq \frac{a \pi }{4} \cos (\phi_2 -\phi_3) |w| \theta(1+\tilde{s}_2)\theta (-\tilde{s}_2)\int_{-\frac{1}{2}-i \infty}^{-\frac{1}{2}+i \infty}\frac{d \omega}{2 \pi i}
\frac{(-2i)\;  \left|\tilde{s}_2\right|^{\omega} }{\omega (\omega+1)}   \left(|w|^{a \tilde{\gamma}_1(\omega)}-1\right),
\eeqn
analytically continued from negative to positive values of $\tilde{s}_2$.
Again the discontinuity is given by a convergent integral at negative $\tilde{s}_2$, but its continuation $|\tilde{s}_2| \rightarrow e^{-i\pi}|\tilde{s}_2|$   requires  a special  consideration.
The expressions (\ref{DeltaAs2}) and (\ref{DeltaBs2}) for discontinuities of the analytic continuations along the paths $\mathbf{A}$ and $\mathbf{B}$ are in an agreement with corresponding expressions for $h_{k}(\sigma)$~(see~(\ref{contAB})). Note, that in both cases the multiplier $\sin \pi \omega$ in the denominator of the integrand is canceled. The factor $1/\sin \pi \omega$ can be considered as the usual signature factor in the Regge formulas allowing to obtain in the physical region of the $t_2$-channel the representation of amplitudes in terms of the Fourier sum of the partial wave contributions with positive integer values of $\omega$.
In our case, however, we have a problem of returning to this Fourier-series representation because the $t$-channel partial waves have the additional pole at $\omega=0$ and essential singularities at the points $\omega=1,2,3,...$ from the expansion of the exponent $|w|^{a \tilde{\gamma}_1 (\omega)}$ in the series in powers of $a$. Note, that for amplitudes with the color-singlet quantum numbers in the $t$-channel such essential singularities are absent because the anomalous dimensions $\gamma$ in this case do not have any poles at $\omega >0$. The simplest example is the one-loop anomalous dimension in the $\mathcal{N}=4$ SYM
\beqn
\gamma^{singlet} =a ( \psi(1)-\psi(\omega)).
\eeqn
Although the infrared divergencies in $\mathcal{N}=4$ SYM  could lead to the absence of the Fourier sum  expansion in the physical region,  the difference in the analytic properties of the $t$-channel partial waves in the $\omega$-plane for the color singlet and adjoint representations looks strange. The question arises: whether or not one can construct the operator product expansion in the collinear limit in such a way, that the essential singularities of the partial waves in the corresponding semi-planes of the $\omega$-plane would be absent.  We see  only one possibility of answering positively  this question, namely,  that the renormalization could not be multiplicative and there should  be  at least two operators having different anomalous dimension, which give comparable contributions in the collinear limit. To discuss this possibility, let us consider the dispersion representation for $R_{OPE}$
\beqn
R_{OPE}(\tilde{s}_2)= \int_{-\infty}^{-1} \frac{d \tilde{s}^{'}_2}{\pi (\tilde{s}^{'}_2-\tilde{s}_2)}\frac{\Delta R^{A}\left(-|\tilde{s}^{'}_2|\right)}{2i}
+\int_{-1}^{0} \frac{d \tilde{s}^{'}_2}{\pi (\tilde{s}^{'}_2-\tilde{s}_2)}\frac{\Delta R^{B}\left(-|\tilde{s}^{'}_2|\right)}{2i},
\eeqn
where
\beqn
\Delta R (-|\tilde{s}_2|)=R(e^{-i\pi} |\tilde{s}_2|)-R(e^{i\pi }|\tilde{s}_2|)
\eeqn
is given by expression
\beqn\label{fABomega}
\frac{\Delta R^{A,B}}{2i }=\frac{a \pi }{4} \cos (\phi_2 -\phi_3)|w| \int_{-\frac{1}{2}-i\infty}^{-\frac{1}{2}+i\infty} \frac{d \omega}{2\pi i }|\tilde{s}_2|^\omega f_{\omega}^{A,B}(w).
\eeqn
In (\ref{fABomega})  we consider initially $f_{\omega}^{A,B}(w)$ only in two first orders of the perturbation theory
\beqn
f_{\omega}^{A}(w)+f_{\omega}^{B}(w)=\frac{1}{\omega(\omega+1)} \left(1+a \ln |w| \tilde{\gamma}_1(\omega)\right)
\eeqn

\beqn
f_{\omega}^{A}(w)\simeq \frac{1}{\omega+1}+ a \ln |w| \frac{\tilde{\gamma}^{-}_1 (\omega)-2 \omega-1}{\omega (\omega+1)}, \;\;
f_{\omega}^{B}(w)\simeq \frac{1}{\omega} +a \ln |w| \frac{\tilde{\gamma}^{+}_1 (\omega)+2 \omega+1}{\omega (\omega+1)},
\eeqn
together with
\beqn
 \tilde{\gamma}^{+}_1 (\omega)=\psi(1-\omega)-\psi(1), \;\;\tilde{\gamma}^{-}_1 (\omega)=\psi(\omega+2)-\psi(1),
\eeqn
where we took into account, that the partial wave $f_{\omega}^{A}(w)$  should not have singularities for $\omega >-1/2$ and $f_{\omega}^{B}(w)$  should not have singularities for $\omega <-1/2$  to provide vanishing $\Delta R^{A}$ in the region $|\tilde{s}_2|<1$ as well as vanishing  $\Delta R^{B}$ for $|\tilde{s}_2|>1$.
If we consider an analogy with the deep-inelastic  $e-p$ scattering, the multiplicative renormalization takes place for the partial waves of the structure functions related directly to the imaginary part of the $\gamma^{*}p $ scattering amplitudes.
In this case the momenta of the structure functions are proportional to the linear combination of matrix elements of the local operators~(for~integer~$\omega$).
The local operators can mix each with others in the course of the renormalization and therefore in a general case they are not renormalized in a multiplicative way.
In the case of the remainder function in the collinear kinematics we also can expect that its discontinuities $\Delta R^{A,B}$ in the $\omega$-plane are  related to linear combinations of  some operators. The comparatively simple situation will be if the number of the relevant operators is finite.
In this case we  can expect that OPE will be valid in the Mandelstam and other physical regions.

However, if we consider the discontinuities  (\ref{DeltaAs2}) and (\ref{DeltaBs2})  for the collinear limit (\ref{ROPEw}) of the remainder function, we obtain a very complicated result.  Namely, $f_{\omega}^{A,B}(|w|)$ entering  (\ref{fABomega})  are given by the expressions
\beqn
f_{\omega}^{A} (|w|)=\int_{-\frac{1}{2}-i\infty}^{-\frac{1}{2}+i\infty} \frac{d \omega'}{2\pi i } \;\frac{1}{\omega'-\omega}\;\frac{1}{\omega'(\omega'+1)} \;e^{a \ln |w| \tilde{\gamma}_1(\omega')}
\eeqn
for $ \Re \;( \omega) >-1/2$ and
\beqn
f_{\omega}^{B} (|w|)=\int_{-\frac{1}{2}-i\infty}^{-\frac{1}{2}+i\infty} \frac{d \omega'}{2\pi i } \;\frac{1}{\omega'-\omega}\;\frac{1}{\omega'(\omega'+1)} \;e^{a \ln |w| \tilde{\gamma}_1(\omega')}
\eeqn
for $ \Re \;( \omega) <-1/2$. Their sum is equal to the total partial wave
\beqn\label{fABomega11}
f_{\omega}^{A}(|w|)+f_{\omega}^{B}(|w|)=\frac{e^{a \ln |w| \tilde{\gamma}_1(\omega)}}{\omega(\omega+1)}.
\eeqn
However,  $f_{\omega}^{A}(|w|)$ and $f_{\omega}^{B}(|w|)$ cannot be written as a finite sum of the exponential terms $~e^{a \ln |w| \tilde{\gamma}_i(\omega)}$.
Moreover, the simple renormalization properties will be absent in all Mandelstam regions obtained by the analytic continuation through the corresponding cuts in the $\tilde{s}_2$-plane.
 In our opinion this reflects some weakness of the simple exponentiation in the AGMSV remainder function.   On the other hand, one can try to make an assumption, that the partial wave contains a sum of two exponents
\beqn\label{fABomegaTwo}
f_{\omega}(|w|) =\frac{1}{\omega(\omega+1)}\left(
e^{a \ln |w| \tilde{\gamma}^{+}_1(\omega)}+ e^{a \ln |w| \tilde{\gamma}^{-}_1(\omega)}
\right),
\eeqn
 where the anomalous dimensions are given below
\beqn
\tilde{\gamma}^{+}_1(\omega)=\psi (2+\omega)-\psi(1),\;\;
\tilde{\gamma}^{+}_1(\omega)=\psi (1-\omega)-\psi(1), \;\;
\tilde{\gamma}_1(\omega)=\tilde{\gamma}^{+}_1(\omega)+\tilde{\gamma}^{-}_1(\omega)
\eeqn
and contain the poles only in the left or in the right semiplanes of the $\omega$-plane. At two loops the expressions in (\ref{fABomega11}) and (\ref{fABomegaTwo}) coincide. At one loop they differ by a a factor of $2$, but in the remainder function the one loop contribution should be subtracted. In principle we can subtract from (\ref{fABomegaTwo})  the term $1/\omega/(\omega+1)$ with the anomalous dimension equal to zero to reproduce the one-loop result.

For the ansatz (\ref{fABomegaTwo}) one can easily find the functions $f_{\omega}^{A}(|w|)$ and $f_{\omega}^{B}(|w|)$  for the discontinuities $A$ and $B$
\beqn\label{modfABomega}
&& f_{\omega}^{A}(|w|) =\frac{1}{\omega (\omega+1)} \left(e^{a \ln |w| \tilde{\gamma}^{+}_1(\omega)}-(1+2\omega)e^{a \ln |w|}\right), \\
&& f_{\omega}^{B}(|w|) =\frac{1}{\omega (\omega+1)} \left(e^{a \ln |w| \tilde{\gamma}^{-}_1(\omega)}+(1+2\omega)e^{a \ln |w|}\right).
\eeqn
The function $f_{\omega}^{A}(|w|) $ is analytic for $\omega >-1/2$ and the function $f_{\omega}^{B}(|w|) $ is analytic for $\omega<-1/2$ in accordance  to the fact, that $\Delta R^{A, B} $ are zero for $-1<\tilde{s}_2<0$  and $\tilde{s}_2 <-1$, respectively.

Thus for the ansatz (\ref{fABomegaTwo}) we reproduce  correctly the expression for the AGMSV remainder function $R_{OPE}$ of Alday et al.~\cite{OPEmalda} at two loops, but at higher loops the predictions are different. Note, that the expressions $f_{\omega}^A(|w|)$ and $f_{\omega}^B(|w|)$  contain the exponents for which the anomalous dimension is the constant $a$.  In the framework of the AdS/CFT correspondence the anomalous dimension is related to the energies of the string states in the Anti-de-Sitter space. Therefore, there should exist a  string state in the adjoint representation, for which the energy does not depend on the angular momentum $\omega$.  Such a state should not have an inner structure and could be a gluon, which can be considered as an elementary particle,  at least in our  approximation.

Note, that both of the ansatz (\ref{fABomega11})  and (\ref{fABomegaTwo}) are in agreement with the Regge asymptotics in the leading double-logarithmic approximation, as it was demonstrated in section~\ref{sec:45text}.  In that section we considered separate contributions from $\gamma_1^{+}(p)$ and $\gamma_1^{-}(p)$ in (\ref{gammaminus}) to the leading logarithmic accuracy in the Mandelstam region. We found that the leading order BFKL result is fully reproduced if one takes into account only powers of $\gamma_1^{-}(p)$ in the AGMSV remainder function $R_{OPE}$ in (\ref{ROPEexp}) up to five loops. $\gamma_1^{+}(p)$ contributes only at next-to-leading logarithmic level, which is  not captured by the LLA BFKL analysis discussed in this study.  This presents another argument in favor of the separate exponentiation of   $\gamma_1^{+}(p)$ and $\gamma_1^{-}(p)$   in  (\ref{fABomegaTwo}).  To resolve the ambiguity in the different exponentiation prescriptions it is needed to calculate the Mandelstam cut contribution in the next-to-leading approximation, which will be hopefully  obtained  in the near future.  In the conclusion we want to stress, that the ansatz (\ref{fABomegaTwo})  is the simplest one, which gives the finite superposition of the exponential terms $\propto e^{a \ln |w| \tilde{\gamma}_i(\omega)}$ for the discontinuities $\Delta^{A}$ and $\Delta^{B}$. Therefore its verification in the next-to-leading BFKL calculation would be important.

\section{Conclusions and discussions}

In the present paper we studied the collinear and Regge limits of the $2\to 4$ MHV amplitude. In particular we considered  the   analytic structure of the remainder function in the collinear kinematics proposed by Alday, Gaiotto, Maldacena, Sever and Vieira~(AGMSV) and continued it analytically to the Mandelstam region. After the continuation, the AGMSV expression  in the Regge limit reproduces the  BFKL results in the double-logarithmic approximation  up to five-loop level. However, we also note that all of the contributions reproducing the known BFKL expressions can be obtained from only one piece of the anomalous dimension $\gamma_1(p)$~(see (\ref{gamma})) present in the AGMSV formula. This piece $\gamma_{1}^{-}(p)$, defined by (\ref{gammaminus}), has singularities only in the lower semiplane of the complex $p$-plane. This translates into the right singularities in the complex angular momentum plane as discussed in section~\ref{sec:reggeinter}.

 In the Regge theory one can expect a clear separation between the right and the left singularities in the complex angular momentum plane, suggesting a non-multiplicative renormalization of the remainder function in the Euclidean region of the  collinear kinematics.  In other words there could be at least two operators having different anomalous dimensions. This  gives the same result as a simple one-operator renormalization at two loops.  The difference between the simple renormalization of the AGMSV expression and the two-operator renormalization suggested in section~\ref{sec:reggeinter} appears already at $3$ loops  and can be verified  only by taking into account next-to-leading corrections to the BFKL eigenvalue in the adjoint color representation. These can be extracted from the NLO BFKL Kernel calculated by Fadin and Fiore~\cite{Fadin:2004zq,Fadin:2005zj} and will  hopefully be found in the near future.

\section{Acknowledgments}

We thank L.~D.~Faddeev, V.~S.~Fadin, G.P. Korchemsky,  E.~M.~Levin, J.~Maldacena, A.~Sabio~Vera, V.~Schomerus, A.~Sever, M.~Spradlin, Chung-I~Tan, C.~Vergu, P.~Vieira and A.~Volovich   for helpful discussions.

 \newpage

\appendix

\setcounter{equation}{0}

\renewcommand{\theequation}{A.\arabic{equation}}
\section{Double Leading Logarithmic Approximation~(DLLA)}\label{app:bessel}

We consider the Double Leading Logarithmic Approximation~(DLLA) for the remainder function calculated in the BFKL approach. We start with the remainder function in the Leading Logarithmic Approximation~(LLA) given by (\ref{corLLA}) and (\ref{LLA}), where we omit all terms subleading in the logarithm of the energy $\ln s_2 \simeq -\ln(1-u_1)$. This expression was calculated in the multi-Regge kinematics given by (\ref{multicross}). Imposing an additional kinematic constraint, that corresponds to the collinear limit (\ref{collmulticross}) we expand (\ref{LLA}) in powers of $|w|$ in accordance with (\ref{collw}). The leading contribution comes from the conformal spin $n=\pm 1$,  and the only the first term in $E_{\nu,n}$ of (\ref{Enun}) has the relevant  poles. We can approximate
\beqn
E_{\nu,n} \simeq E_{\nu,1} \simeq -\frac{1}{2}\frac{1}{\nu^2 +\frac{1}{4}}
\eeqn
in the double logarithmic approximation and write
\beqn\label{DLLAapp}
&&R^{DLLA}_{BFKL}\simeq 1 -i\frac{a}{2} \left(\frac{w}{w^*}+\frac{w^*}{w} \right) \sum_{k=1}^{\infty} \frac{ a^k \ln^{k}(1-u_1) }{k!} \int_{-\infty}^{\infty}\frac{ d\nu \; |w|^{2i \nu}}{\nu^2+\frac{1}{4}} E^k_{\nu,1} \\
&&
\simeq1 -i a \cos (\phi_2-\phi_3) \sum_{k=1}^{\infty} \frac{(-1)^k a^k 2^{-k} \ln^{k}(1-u_1)}{k!} \int_{-\infty}^{\infty}\frac{ d\nu  \; |w|^{2i \nu}}{\left(\nu^2+\frac{1}{4}\right)^{k+1}} \hspace{1cm}  \\
&&  \simeq 1-\; i2\pi a \cos (\phi_2-\phi_3)\;|w| \sum_{k=1}^{\infty} \frac{ a^k \ln^k |w| \ln^{k}(1-u_1)}{(k!)^2} \\
&&= 1+
 i2\pi a \cos (\phi_2-\phi_3) \;|w| \left(1-I_0\left(2 \sqrt{a \ln |w| \ln(1-u_1)}\right) \right),
\eeqn
where $I_0(z)$ is the modified Bessel function.

The contribution to the real part of the NLLA remainder function comes from several terms in the dispersion-like  relation in (\ref{disp}). Expanding (\ref{disp})  to the second and the third order in powers of $a$ we obtain
\beqn\label{expanDisp2app}
a^2 R^{(2)}-\frac{\pi^2 \delta^2}{2}=-\frac{\pi^2 \omega^2_{ab}}{2}
+i\frac{a^2}{2}\frac{\partial^2}{\partial a^2} \left( \int_{-i\infty}^{i\infty} \frac{d \omega}{2\pi i} f(\omega) e^{-i\pi \omega} |1-u_1|^{-\omega}\right)
\eeqn
and
\beqn\label{expanDisp3app}
a^3 R^{(3)}+i \pi \delta a^2 R^{(2)}-i\frac{\pi^3 \delta^3}{6}=i\frac{a^3}{6}\frac{\partial^3}{\partial a^3} \left( \int_{-i\infty}^{i\infty} \frac{d \omega}{2\pi i} f(\omega) e^{-i\pi \omega} |1-u_1|^{-\omega}\right).
\eeqn
We are interested only  in the leading logarithmic~(LLA) and the real part of the next-to-leading~(NLLA) in the logarithm $\ln(1-u_1)$ contributions. Thus we can omit all subleading terms in (\ref{expanDisp2app}) and (\ref{expanDisp3app})   as follows
\beqn\label{expanDisp2appNLLA}
a^2 R_{BFKL}^{(2)}-\frac{\pi^2 \delta^2}{2}=-\frac{\pi^2 \omega^2_{ab}}{2}
+i\frac{a^2}{2}\frac{\partial^2}{\partial a^2} \left( \int_{-i\infty}^{i\infty} \frac{d \omega}{2\pi i} f^{LLA}(\omega) e^{-i\pi \omega} |1-u_1|^{-\omega}\right)
\eeqn
and
\beqn\label{expanDisp3appNLLA}
a^3 R_{BFKL}^{(3)}+i \pi \delta a^2 R_{BFKL}^{(2)}=i\frac{a^3}{6}\frac{\partial^3}{\partial a^3} \left( \int_{-i\infty}^{i\infty} \frac{d \omega}{2\pi i} f^{LLA}(\omega) e^{-i\pi \omega} |1-u_1|^{-\omega}\right).
\eeqn

The integral in RHS of (\ref{expanDisp2appNLLA}) and (\ref{expanDisp3appNLLA}) is related to $\Delta$ defined in (\ref{LLA}) by
\beqn
\Delta^{LLA}_{2\to4}= \int_{-i\infty}^{i\infty} \frac{d \omega}{2\pi i} f^{LLA}(\omega)  |1-u_1|^{-\omega}
\eeqn
for  $f^{LLA}(\omega)$ given by (\ref{fomegaLLA}), and was calculated to the second~\cite{BLS2} and the third order~\cite{LP2} in $a$.
 The phase of its integrand  $ e^{-i\pi \omega}$ can be accounted for by making a substitution $\ln(1-u_1) \to \ln(1-u_1)+i\pi$ in the final result.
At two loops we have a full cancellation between real NLLA contributions coming from the integral, $\delta$ and $\omega_{ab}$. This is not the case at three loops, where the real part of the NLLA remainder function was calculated in ref.~\cite{LP2} and reads

\beqn\label{R3NLLAReapp}
&&  R_{BFKL}^{(3)\;LLA}=i\Delta_{2\to4}^{(3)} /a^3=i\pi \frac{1}{4} \ln^2(1-u_1)\left(
 \ln|w|^2\ln^2|1+w|^2-\frac{2}{3}\ln^3|1+w|^2\right. \hspace{1cm}\;\;\;
\\
&&\left.-\frac{1}{4}\ln^2|w|^2 \ln|1+w|^2+\frac{1}{2} \ln|w|^2 \left(\text{Li}_2(-w)+\text{Li}_2(-w^*)\right)
 - \text{Li}_3(-w)-\text{Li}_3(-w^*)\right). \nonumber
\eeqn

Using dispersion-like  relation (\ref{disp}), it is possible to write a general relation between  the LLA  and the real part of NLLA  remainder function at an arbitrary number of loops
\beqn\label{dispReNLLAapp}
\Re\left(R^{NLLA}_{BFKL}\right)= 1+i\pi \frac{\partial R_{BFKL}^{(LLA)}}{\partial \ln(1-u_1) } -i\pi \delta R_{BFKL}^{(LLA)}+\frac{\pi^2}{2 }\left(\delta^2-\omega^2_{ab}\right),
\eeqn
where $\delta$ and $\omega_{ab}$ are given by (\ref{deltaomega}), and the integral representation for $R_{BFKL}^{(LLA)}$ is known and can be analytically calculated to any order.

In the double logarithmic approximation~(DLLA), where  both $1-u_1$ and $|w|$ are small, the relation (\ref{dispReNLLAapp}) can be explicitly calculated as follows. First, we find
\beqn\label{t3app}
\frac{\pi^2}{2 }\left(\delta^2-\omega^2_{ab}\right)=\frac{a^2 \pi^2}{2} \ln \left|1+w \right|^2\ln \left|1+\frac{1}{w} \right|^2
\simeq -2\pi^2 a^2 \cos(\phi_2-\phi_3) |w| \ln |w|,
\eeqn
next, using (\ref{DLLAapp}) we readily obtain
\beqn\label{t2app}
-i\pi \delta R_{BFKL}^{(LLA)}\simeq   4 \pi^2 a \cos (\phi_2-\phi_3) \;|w|\; \ln|w| \left(1-I_0\left(2 \sqrt{a \ln |w| \ln(1-u_1)}\right) \right).
\eeqn
Finally, the first term in RHS of (\ref{dispReNLLAapp}) can be calculated replacing $\ln^k(1-u_1)$ in (\ref{DLLAapp}) by $  i\pi k \ln^{k-1} (1-u_1))$, because of the phase of the integrand $e^{-i\pi \omega}$ in the dispersion relation (\ref{disp}) generates terms with $\ln (1-u_1) \to \ln (1-u_1) +i\pi $ and we are interested only in the next-to-leading contributions
\beqn\label{t1app}
&& i\pi \frac{\partial R_{BFKL}^{(LLA)}}{\partial \ln(1-u_1) }= a \; 2\pi^2 \cos (\phi_2-\phi_3)\;|w| \sum_{k=1}^{\infty} \frac{ a^k \ln^k |w| k\ln^{k-1}(1-u_1)}{(k!)^2} \\
&&=
 2\pi^2 a^{3/2} \cos (\phi_2-\phi_3)\; |w| \;\ln|w| \frac{I_1\left(2 \sqrt{a \ln |w| \ln(1-u_1)}\right)}{\ln(1-u_1)}. \nonumber
\eeqn
 Plugging (\ref{t3app}), (\ref{t2app}) and (\ref{t1app}) in (\ref{dispReNLLAapp}) we get in the double logarithmic approximation the real part of the contribution subleading in $\ln(1-u_1)$
\beqn\label{ReNLLAapp}
&&\Re\left(R^{DNLLA}_{BFKL}\right)\simeq1+ 2\pi^2 a^{3/2} \cos (\phi_2-\phi_3)\; |w| \;\ln|w| \frac{I_1\left(2 \sqrt{a \ln |w| \ln(1-u_1)}\right)}{\ln(1-u_1)} \\
&& +  4 \pi^2 a \cos (\phi_2-\phi_3) \;|w|\; \ln|w| \left(1-I_0\left(2 \sqrt{a \ln |w| \ln(1-u_1)}\right) \right)
-2\pi^2 a^2 \cos(\phi_2-\phi_3) |w| \ln |w|. \nonumber
\eeqn


\setcounter{equation}{0}

\renewcommand{\theequation}{B.\arabic{equation}}
\section{The AGMSV remainder function at three loops}\label{app:h2}

In this section we present   details of the calculation of the three loop contribution to the remainder function in the collinear limit given by (\ref{ROPEexp}). At three loops  it reads
\beqn\label{Rl3app}
R^{(3)}_{OPE}\simeq \cos \phi \;e^{-\tau}  \frac{\tau^{2}}{2}\int  c^{0}(p)\gamma^{2}_1 (p)   e^{ip\sigma} dp,
\eeqn
where
\beqn\label{coapp}
c^{0}(p) =\frac{2}{1+p^2}\frac{1}{\cos \frac{p \pi}{2}}
\eeqn
and
\beqn\label{gamma1app}
\gamma_1(p)=\psi\left( \frac{3}{2}+i\frac{p}{2}\right)+\psi\left( \frac{3}{2}-i\frac{p}{2}\right)-2\psi\left(1\right).
\eeqn
In order to find $R^{(3)}_{OPE}$ we need to calculate
\beqn
h_2(\sigma)=\int_{-\infty}^{\infty}  c^{0}(p) \gamma^2_1(p) e^{ip \sigma} dp
\eeqn
defined in (\ref{hk}).
At two loops the remainder function $R^{(2)}_{OPE}$ in the collinear limit was found in ref.~\cite{OPEmalda} and the relevant integral reads
\beqn
h_1(\sigma)=\int_{-\infty}^{\infty} c^{0}(p) \gamma_1(p)  e^{ip \sigma} dp,
\eeqn
where
\beqn\label{h1}
h_1(\sigma)=-2\cosh \sigma \left(2 \ln \left( 1+e^{2\sigma}\right)\ln \left( 1+e^{-2\sigma}\right)-4\ln \left(2 \cosh \sigma\right)\right)
-8 \sigma \sinh \sigma.
\eeqn
From (\ref{h1}) we see that the most complicated term is given by $\cosh \sigma \ln^2 \left(2 \cosh \sigma\right)$. At three loops  it is natural to expect  $\cosh \sigma \ln^3 \left(2 \cosh \sigma\right)$, but this function diverges at $\sigma \to \pm \infty$ and need to be cured by $\sigma^3 \sinh \sigma$ term. Other way to cure the divergency is to introduce the regularization parameter $p \to p- i \epsilon$ in the integral
\beqn\label{int1app}
\int_{-\infty}^{\infty} \cosh \sigma \ln^3 \left(2 \cosh \sigma\right) e^{-i(p -i\epsilon)\sigma} d \sigma=
\frac{\partial^3}{\partial a^3}\frac{1}{2}\int_0^{\infty}\left(z+\frac{1}{z}\right)^a z^{-i(p -i\epsilon)-1} dz|_{a=0},
\eeqn
where $z=e^{\sigma}$. The last integral in (\ref{int1app}) gives the Euler Beta function and its higher derivatives that give the polygamma functions. In an analogous way we calculate the term $\sigma^3 \sinh \sigma$ and obtain the final expression with no $\epsilon$ dependence
\beqn\label{coshlog2app}
&& 8 \left(\cosh \sigma \ln^3 \left(2 \cosh \sigma\right)-\sigma^3 \sinh \sigma\right)
\\
&&
=\int_{-\infty}^{\infty} c^{0}(p)
\left(-6 \gamma_{1}(p)+\frac{3}{2}\gamma^2_{1}(p)+\frac{3}{2}\tilde{\gamma}_1(p)+\frac{48}{(1+p^2)^2}-\frac{24}{1+p^2}-\frac{\pi^2}{2}\right) e^{ip\sigma} dp,\;\;\; \nonumber
\eeqn
where
\beqn\label{gamma1tildeapp}
\tilde{\gamma}_1(p)=\psi'\left( \frac{3}{2}+i\frac{p}{2}\right)+\psi'\left( \frac{3}{2}-i\frac{p}{2}\right)-2\psi'\left(1\right).
\eeqn
and the functions $\gamma_1(p)$ and $c^{0}(p)$ are given by (\ref{gamma1app}) and (\ref{coapp}) respectively.
The first term on RHS of (\ref{coshlog2app}) is proportional to $h_1(p)$ and the last term is known from ref.~\cite{OPEmalda}
\beqn
h_0(\sigma)=\int_{-\infty}^{\infty}c^{0}(p)e^{ip\sigma}dp =4 \cosh \sigma \ln (2 \cosh \sigma)-4\sigma \sinh \sigma.
\eeqn
The rest of the terms in (\ref{gamma1tildeapp}) are calculated using the Cauchy theorem.
For simplicity we consider only the case of positive $\sigma$ closing the integration contour in the upper semiplane. The terms in (\ref{gamma1tildeapp}) have a higher order pole at $p=i$ and all other poles are simple poles. Thus we can readily write
\beqn\label{power3app}
&&I_1=\int_{-\infty}^{\infty}c^{0}(p)\frac{1}{(1+p^2)^2} e^{ip\sigma} dp
\;\;\;  \\
&&
=\frac{ e^{-\sigma}}{2}\left(\frac{5}{2}+\frac{\pi^2}{8}+3 \sigma +\frac{\pi^2 \sigma}{12}+\frac{3 \sigma^2}{2}+\frac{\sigma^3}{3}\right)
-\sum^{\infty}_{n=1}\frac{(-1)^n  e^{-\sigma(1+2n)}}{8 n^3 (1+n)^3}
\;\;\; \nonumber \\
&&
=\frac{ e^{-\sigma}}{2}\left(\frac{\pi^2}{8}+\frac{\pi^2 \sigma}{12}+\frac{3 \sigma^2}{2} +\frac{\sigma^3}{3}\right)+\frac{3}{8}h_{0}
-\frac{3}{4}\sinh \sigma \;\text{Li}_2(-e^{-2\sigma})
-\frac{1}{4}\cosh \sigma \;\text{Li}_3(-e^{-2\sigma}), \nonumber
\eeqn
where the first term on  RHS comes from the pole at $p=i$.

The transform of $c^{0}(p)/(1+p^2)$  can be obtained directly from (\ref{power3app}) differentiating it twice  with respect to $\sigma$
\beqn
I_2=\int_{-\infty}^{\infty}c^{0}(p)\frac{1}{1+p^2} e^{ip\sigma} dp= -\frac{d^2 I_1 }{d \sigma^2}+I_1=\frac{\pi^2 e^{-\sigma}}{12}+ \sigma^2 e^{-\sigma}+ \frac{h_{0}}{2}-\sinh \sigma \; \text{Li}_2(-e^{-2\sigma})\;\;\;
\eeqn

Finally we calculate the last missing contribution in (\ref{coshlog2app})

\beqn
&&\int_{-\infty}^{\infty}c^{0}(p)\tilde{\gamma}_1(p) e^{ip\sigma} dp  =-6  e^{-\sigma}-4\sigma e^{-\sigma}
-\sum_{n=1}^{\infty}\frac{(-1)^n 2 e^{-\sigma(1+2n)}}{n (1+n)}\left(\frac{1+2\sigma}{n(n+1)}
\zeta_2+\frac{4\sigma}{n+1}+2\sigma^3 \nonumber
\right)
\\
&&=-\frac{\pi^2 e^{-\sigma}}{3}-4 \sigma^2 e^{-\sigma}+2 \cosh \sigma
\left( -\frac{\pi^2}{3}-4\sigma^3-4 \ln (2\cosh \sigma)+\frac{\pi^2}{3} \ln (2\cosh \sigma)\right.  \\
&& \left.+4\sigma^2 \ln (2\cosh \sigma)-4 \text{Li}_2(-e^{-2\sigma})\right)+2\pi \sinh \sigma \left(2\sigma +\text{Li}_2(-e^{-2\sigma})\right)
\nonumber
\eeqn

 This allows us to find the integral in the three loop expression (\ref{Rl3app})
\beqn \label{h2app}
&& h_2(\sigma)=\int_{-\infty}^{\infty}c^{0}(p)\gamma^2_1(p) e^{ip\sigma} dp=-\frac{\pi^2}{3} e^{-\sigma }-4 e^{-\sigma }   \sigma ^2-\frac{2}{3} \pi ^2 \sigma  \cosh \sigma
\\
&&+16  \sigma ^2 \cosh \sigma+\frac{8}{3}  \sigma ^3 \cosh \sigma+24  \cosh \sigma \ln (2 \cosh \sigma) \nonumber
+\frac{2}{3} \pi ^2 \cosh \sigma \ln (2 \cosh \sigma)
\\
&&
-8  \sigma ^2 \cosh \sigma \ln (2 \cosh \sigma)
-16  \cosh \sigma \ln^2 (2 \cosh \sigma) \nonumber
+\frac{16}{3}   \cosh \sigma \ln^3 (2 \cosh \sigma)
\\
&&
+8  \sigma  \cosh \sigma \;\text{Li}_2(-e^{-2\sigma})+8  \cosh \sigma \; \text{Li}_3(-e^{-2\sigma}) \nonumber
-24  \sigma  \sinh \sigma
+4  \sinh \sigma \; \text{Li}_2(-e^{-2\sigma})  \nonumber
\eeqn
 The expression in (\ref{h2app}) vanishes at $\sigma \to \infty$ and is symmetric under $\sigma \to -\sigma$.


\setcounter{equation}{0}

\renewcommand{\theequation}{C.\arabic{equation}}

\section{Analytic continuation}\label{app:cont}

In this section we perform the analytic continuation of the AGMSV remainder function $R_{OPE}$ in (\ref{ROPEexp}). The analytic continuation ($u_2$, $u_3$ are fixed and  $u_1=|u_1| e^{-i2\pi}$), which was used~\cite{LP1,LP2} to extract (\ref{R2LLABFKL}) and (\ref{R2NLLABFKL}) from the GSVV remainder function, is not applicable here. This is because the collinear function $R_{OPE}$  was obtained under  assumption of the finiteness of the cosine factor
\beqn
\cos \phi =\frac{u_1 +u_2 +u_3-1}{2 \sqrt{u_1 u_2 u_3}},
\eeqn
which diverges when we cross a point $u_1=|u_1| e^{-i\pi}$ in the multi-Regge kinematics given by (\ref{multicross}). At this point the numerator becomes of the order of unity, while the denominator is small. Using the parametrization of the dual conformal cross ratios introduced in ref.~\cite{OPEmalda} and given by (\ref{sigmatauphi}) we find
 \beqn
\frac{u_1}{u_3}=e^{2\sigma},
\eeqn
so that the path of the continuation in the complex $\sigma$-plane is just a shift of $\sigma$, namely
\beqn
\sigma \Rightarrow \sigma -i\pi.
\eeqn
We call this path of the  analytic  continuation- \emph{the path} $\mathbf{A}$.

 It was argued in section~\ref{sec:collregge}, that one can smoothly deform the path of the continuation to make it compatible with the collinear limit by taking into account the relation
\beqn
u_3 \to 1- u_1
\eeqn
for $u_2 \to 0$. This makes the numerator of $\cos \phi$ to be of the order of its denominator at $u_1=|u_1| e^{-i\pi}$ and thus to be compatible with the basic assumptions of the collinear expansion in $\tau$.  We can define

\beqn
e^{2\sigma} \simeq \frac{u_1}{1-u_1},\;\; \sigma \simeq \frac{1}{2}\ln \frac{u_1}{1-u_1}
\eeqn
for $u_1=|u_1| e^{-i\pi}$ as the analytic continuation along \emph{the path} $\mathbf{B}$. In the continuation with the path  $\mathbf{B}$ the cross ratio $u_3$ is not fixed anymore and  possesses a non-trivial phase as $u_1$ rotates around the origin.  The paths  $\mathbf{A}$ and $\mathbf{B}$ in the complex $\sigma$-plane are illustrated in Fig.~\ref{fig:cuts}. Here list the analytic continuation along the path $\mathbf{B}$ of the functions relevant at two and three loops. By the words ``Regge limit'' over the arrows we mean the multi-Regge  and collinear kinematics (\ref{collmulticross}) for which $\sigma \simeq -1/2 \ln(1-u_1) \to +\infty$.

\beqn
\sigma \simeq \frac{1}{2}\ln u_1-\frac{1}{2}\ln (1-u_1)  \Rightarrow -i\pi+ \sigma  \overbrace{\longrightarrow}^{\text{Regge limit}}
-i\pi -\frac{1}{2}\ln(1-u_1)
\eeqn

\beqn
e^{\sigma}\simeq\sqrt{\frac{u_1}{1-u_1}} \Rightarrow -e^{\sigma} \overbrace{\longrightarrow}^{\text{Regge limit}} \frac{-1}{\sqrt{1-u_1}}, \;\;\;  e^{-\sigma}\simeq \sqrt{\frac{1-u_1}{u_1}} \Rightarrow -e^{-\sigma} \overbrace{\longrightarrow}^{\text{Regge limit}} 0, \;\;\;
\eeqn

\beqn
 \cosh \sigma  \Rightarrow -\cosh \sigma \overbrace{\longrightarrow}^{\text{Regge limit}} \frac{-1}{2\sqrt{1-u_1}}, \;  \sinh \sigma  \Rightarrow -\sinh \sigma \overbrace{\longrightarrow}^{\text{Regge limit}} \frac{-1}{2\sqrt{1-u_1}}
\eeqn

\beqn
\ln (2 \cosh \sigma)\simeq \ln \left(\frac{1}{ \sqrt{u_1 (1-u_1)}}\right)\Rightarrow  i\pi  +\ln (2 \cosh \sigma)
\overbrace{\longrightarrow}^{\text{Regge limit}} i\pi -\frac{1}{2}\ln(1-u_1)
\eeqn

\beqn
&& \text{Li}_2(-e^{-2\sigma})=\text{Li}_2\left(\frac{u_1-1}{u_1}\right)=
-\int_0^{\frac{u_1-1}{u_1}}\frac{dt}{t}\ln (1-t)  \Rightarrow \text{Li}_2(-e^{-2\sigma}) -i2\pi \int_1^\frac{u_1-1}{u_1} \frac{dt}{t} \hspace{1cm}\nonumber
\\
&&=
 \text{Li}_2(-e^{-2\sigma}) -i2\pi (-2\sigma+i\pi)\overbrace{\longrightarrow}^{\text{Regge limit}} -i2\pi (\ln(1-u_1)+i\pi)
\eeqn

\beqn
&&\text{Li}_3(-e^{-2\sigma})  \Rightarrow \text{Li}_3(-e^{-2\sigma}) -i2\pi \int_1^\frac{u_1-1}{u_1} \frac{dt}{t} \int^t_1 \frac{dt'}{t'}
=
 \text{Li}_3(-e^{-2\sigma}) -i2\pi \frac{(-2\sigma+i\pi)^2}{2} \nonumber
 \\
&&
\overbrace{\longrightarrow}^{\text{Regge limit}} -i2\pi \frac{(\ln(1-u_1)+i\pi)^2}{2}
\eeqn

Having the list of  all necessary functions  continued along the path $\mathbf{B}$, we can readily write the analytic continuation of $h_k(\sigma)$ defined by (\ref{hk}). Note that the  general structure of $h_k(\sigma)$ after the  analytic continuation is
\beqn
h_k (\sigma) \Rightarrow - h_k (\sigma)+ \Delta_k (\sigma),
\eeqn
where $h_k(\sigma)$ changes the sign and receives an additive function. The change of the sign is related to the fact that $\cos \phi$ also changes the sign
\beqn
\cos \phi =\frac{u_1 +u_2 +u_3-1}{2 \sqrt{u_1 u_2 u_3}}\Rightarrow -\cos \phi \overbrace{\longrightarrow}^{\text{Regge limit}}
-\frac{u_1 +u_2 +u_3-1}{2 \sqrt{ u_2 u_3}}=\cos \left(\phi_2 -\phi_3\right)
\eeqn
so that the product of $\cos \phi$ and $h_k (\sigma)$ that appears in the AGMSV remainder function in (\ref{ROPEexp}) has the same sign and gets an additive function after the analytic  continuation. We start with $h_0(\sigma)$ in (\ref{h0}), which corresponds to one loop, i.e. the BDS amplitude
\beqn\label{h0contapp}
h_{0} (\sigma) \Rightarrow -h_{0}(\sigma) -i4\pi e^{\sigma} \overbrace{\longrightarrow}^{\text{Regge limit}} \frac{-i4\pi}{\sqrt{1-u_1}}
\eeqn
and read out of it
\beqn
\Delta_0 (\sigma)=-i4\pi e^{\sigma}.
\eeqn
Next the we consider $h_1(\sigma)$ for the AGMSV remainder function at two loops
\beqn\label{h1contapp}
h_1(\sigma) \Rightarrow -h_{1}(\sigma)-i8\pi e^{\sigma}+i8\pi \cosh \sigma \ln(1+e^{2 \sigma})\overbrace{\longrightarrow}^{\text{Regge limit}}-\frac{i4\pi}{\sqrt{1-u_1}}\ln(1-u_1)-\frac{i8\pi}{\sqrt{1-u_1}}, \;\;\; \;\;
\eeqn
which gives
\beqn
\Delta_1 (\sigma)=-i8\pi e^{\sigma}+i8\pi \cosh \sigma \ln(1+e^{2 \sigma}).
\eeqn

At three loops we have
\beqn\label{h2contapp}
&& h_2(\sigma) \Rightarrow -h_2(\sigma) - i 24 \pi e^{\sigma }+4 i h_1(\sigma) \pi -4 \pi^2 e^{\sigma } +2 \pi^2 h_0(\sigma) + i 24 \pi e^{\sigma }  \sigma \\
&& -4 i \pi h_0(\sigma)  \sigma  +8 \pi^2 e^{\sigma }  \sigma  +\frac{4}{3} i \pi ^3 \cosh \sigma + i8  \pi  \cosh \sigma \text{Li}_2(-e^{-2\sigma})- i 16 \pi  \sigma ^2 \sinh \sigma
\overbrace{\longrightarrow}^{\text{Regge limit}} \nonumber\\
&&
-\frac{i 24 \pi }{\sqrt{1-\text{u1}}}
-\frac{4 \pi ^2}{\sqrt{1-\text{u1}}}
+\frac{i 2 \pi ^3}{3 \sqrt{1-\text{u1}}}-\frac{ i 12 \pi \ln(1-u_1)}{\sqrt{1-\text{u1}}}-\frac{4 \pi ^2 \ln(1-u_1)}{\sqrt{1-\text{u1}}}-\frac{i 2 \pi  \ln^2(1-u_1)}{\sqrt{1-\text{u1}}}, \nonumber
\eeqn
from which we extract
\beqn\label{Delta2}
&& \Delta_{2}(\sigma)=8 i \pi  \cosh  \sigma \text{Li}_2\left(-e^{-2 \sigma }\right)  +16 i \pi  \sigma ^2 \cosh  \sigma -8 i \pi  e^{\sigma } \sigma -4 \pi ^2
   e^{\sigma }-24 i \pi  e^{\sigma }
\\
&&
+\frac{4}{3} i \pi ^3 \cosh  \sigma -16 i \pi  \cosh  \sigma  \ln^2\left(1+e^{2 \sigma }\right)  +16 i \pi
    \sigma \cosh  \sigma  \ln\left(1+e^{2 \sigma }\right)  \nonumber
\\
&&
  +8 \pi ^2 \cosh  \sigma \ln\left(1+e^{2 \sigma }\right) +32 i \pi  \cosh  \sigma \ln
   \left(1+e^{2 \sigma }\right).  \nonumber
\eeqn

The AGMSV remainder function in (\ref{ROPEexp}) at two loops reads
\beqn
R^{(2)}_{OPE} \simeq -\cos \phi \; e^{-\tau} \tau \int_{-\infty}^{\infty} c^{0}(p)  \gamma_1(p) e^{ip\sigma} d p=-\cos \phi \; e^{-\tau}\; \tau \; h_1(\sigma)
\eeqn
and after the analytic continuation along path $\mathbf{B}$ we obtain~(note that $\cos \phi \Rightarrow -\cos \phi$ after the analytic continuation)
\beqn
&& R^{(2)}_{OPE} \Rightarrow \cos \phi \; e^{-\tau}\; \tau \; (-h_{1}(\sigma)-i8\pi e^{\sigma}+i8\pi \cosh \sigma \ln(1+e^{2 \sigma}))
 \\
 && \overbrace{\longrightarrow}^{\text{Regge limit}}_{\sigma \to +\infty}
 -\cos(\phi_2-\phi_3)\frac{\sqrt{u_2}}{2} (\ln 2 -\ln \sqrt{u_2})
 \left(-\frac{i4\pi}{\sqrt{1-u_1}}\ln(1-u_1)-\frac{i8\pi}{\sqrt{1-u_1}}\right) \nonumber
 \\&& = i2\pi\cos(\phi_2-\phi_3) \; |w| \;\left(\ln 2-\ln |w|-\frac{1}{2}\ln(1-u_1)\right)\left(\ln(1-u_1)+2\right),
\eeqn
where we used the definition $|w|^2=u_2/u_3$ and the fact that in the collinear limit we have
\beqn
u_3 \to 1-u_1
\eeqn
 as well as
 \beqn
e^{-\tau} \simeq \frac{\sqrt{u_2}}{2}=\frac{|w| \sqrt{1-u_1}}{2},\;
\tau \simeq\ln 2- \ln \sqrt{u_2}= \ln 2 -\ln |w|-\frac{1}{2} \ln(1-u_1).
\eeqn
Already at this point we see that the terms leading in $\ln|w|$ reproduce the BFKL result. However, in order to find the full agreement we have to include the terms subleading in $\tau$ because their smallness is of the same order as those enhanced by $\ln(1-u_1)$. These we extract from the GSVV expression
\beqn
&& h^{sub}_1(\sigma)=-\frac{2}{3} \pi ^2 \sigma  \cosh \sigma-4 \sigma ^2 \cosh \sigma-\frac{8}{3} \sigma ^3 \cosh \sigma+4 \sigma ^2 \cosh \sigma \ln 2
\\ \nonumber
&&-8 \cosh \sigma \ln(2 \cosh \sigma)+\frac{2}{3} \pi ^2 \cosh \sigma \ln(2 \cosh \sigma)+4 \sigma ^2 \cosh \sigma \ln(2 \cosh \sigma) \\ \nonumber
&&+8 \cosh \sigma  \ln(2 \cosh \sigma)\ln 2+4 \cosh \sigma \ln^2(2 \cosh \sigma)-4 \cosh \sigma  \ln^2(2 \cosh \sigma)\ln 2\\ \nonumber
&&-\frac{4}{3} \cosh \sigma \ln^3(2 \cosh \sigma)+4 \cosh \sigma \text{Li}_3 \left(-e^{-2\sigma}\right)-8 \sigma  \sinh \sigma-8 \sigma   \sinh \sigma \ln 2,
\eeqn
so that   for $\tau \to + \infty$ one can write
\beqn
R^{(2)}_{GSVV}\simeq \cos \phi \; e^{-\tau} \left(-\tau h_1(\sigma)+h^{sub}_1(\sigma)\right)+\mathcal{O}\left(e^{-2\tau}\right).
\eeqn
 After the analytic continuation this gives
 \beqn
  h_1^{sub}(\sigma) \Rightarrow -  h_1^{sub}(\sigma)+\Delta_1^{sub}(\sigma),
 \eeqn
where
\beqn
&& \Delta_1^{sub}(\sigma)=8 i \pi  e^{\sigma }-8 i \pi  e^{\sigma } \ln 2+4 i \pi  \cosh \sigma  \ln ^2\left(1+e^{2 \sigma }\right)
\\
&&
-8 i \pi  \cosh \sigma  \ln \left(1+e^{2 \sigma    }\right)
 +8 i \pi  \ln 2 \cosh \sigma   \ln \left(1+e^{2 \sigma }\right)  \nonumber
\eeqn
and the relevant terms then become
\beqn
&&\cos \phi \; e^{-\tau} h^{sub}_1(\sigma)\Rightarrow -\cos \phi \; e^{-\tau} \left( -  h_1^{sub}(\sigma)+
\Delta_1^{sub}(\sigma) \right)
 \overbrace{\longrightarrow}^{\text{Regge limit}} \nonumber\\
&&
\cos (\phi_2-\phi_3)\frac{\sqrt{u_2}}{2}\left(
 \frac{8 i \pi }{\sqrt{1-u_1}} -\frac{8 i \pi  \ln 2}{\sqrt{1-u_1}}+\frac{4 i \pi  \ln(1-u_1)}{\sqrt{1-u_1}}
 -\frac{4 i \pi  \ln 2 \ln(1-u_1)}{\sqrt{1-u_1}}+\frac{2 i \pi  \ln(1-u_1)^2}{\sqrt{1-u_1}}\right)\nonumber \\
 && =i\pi \cos(\phi_2-\phi_3)\; |w|\;\left(4-4 \ln 2+2 \ln(1-u_1)-2 \ln 2 \ln(1-u_1)+\ln^2(1-u_1) \right).
 \nonumber
\eeqn
Adding this to the AGMSV remainder function we obtain
\beqn
R^{(2)}_{OPE}+\cos \phi \;e^{-\tau} h_1^{sub}(\sigma)\Longrightarrow -i2\pi \cos (\phi_2-\phi_3) \;|w|\;(\ln (1-u_1) \ln |w|+2 \ln |w|
-2),
\eeqn
which  fully reproduces the BFKL result and the subleading corrections extracted from the GSVV expression  given by (\ref{coll2BFKL}) in the collinear limit.

Applying a similar analysis to the AGMSV remainder function at three loops we also reproduce the BFKL result in the collinear and multi-Regge kinematics given by (\ref{collmulticross}) as follows. After the analytic continuation of $h_{2}(\sigma) $ along the path $\mathbf{B}$ we have
\beqn
h_{2}(\sigma)  \Rightarrow -h_{2}(\sigma)+\Delta_{2}(\sigma),
\eeqn
where $\Delta_{2}(\sigma)$ is given by (\ref{Delta2}). Plugging this in the  expression for the remainder function in the collinear limit (\ref{ROPEexp}) we obtain
\beqn\label{R3OPEContapp}
&&R^{(3)}_{OPE}\simeq \cos \phi \; e^{-\tau} \; \frac{\tau^2}{2} \;h_2(\sigma)\Rightarrow -\cos \phi e^{-\tau} \frac{\tau^2}{2} \left( -h_2(\sigma) +\Delta_{2}(\sigma) \right)\overbrace{\longrightarrow}^{\text{Regge limit}} \\
&&
\cos (\phi_2-\phi_3)\frac{\sqrt{u_2}}{2} \frac{\left(-\frac{1}{2}\ln u_2+\ln 2\right)^2}{2} \left(-\frac{i 24 \pi }{\sqrt{1-\text{u1}}} -\frac{4 \pi ^2}{\sqrt{1-\text{u1}}}
+\frac{i 2 \pi ^3}{3 \sqrt{1-\text{u1}}}\right.
\nonumber\\
&&
\left.
-\frac{ i 12 \pi \ln(1-u_1)}{\sqrt{1-\text{u1}}}-\frac{4 \pi ^2 \ln(1-u_1)}{\sqrt{1-\text{u1}}}-\frac{i 2 \pi  \ln^2(1-u_1)}{\sqrt{1-\text{u1}}} \right)\nonumber \\
&& \simeq -\frac{i\pi}{2} \ln^2(1-u_1)\cos (\phi_2-\phi_3)\;|w|\; \ln^2|w|-\pi^2 \ln(1-u_1)\cos (\phi_2-\phi_3)\;|w|\; \ln^2|w|\nonumber\\
&&
-i3\pi \ln(1-u_1)\cos (\phi_2-\phi_3)\;|w|\; \ln^2|w|.
\nonumber
\eeqn
In (\ref{R3OPEContapp}) we omit terms of the order of $\ln^2 |w| $ not enhanced by $\ln (1-u_1)$ because they correspond to the next-to-leading corrections in the logarithm of the energy and are irrelevant for the purpose of the present discussion. The first two terms in RHS of (\ref{R3OPEContapp}) reproduce the BFKL result in the Double Leading Logarithmic Approximation~(DLLA) given by (\ref{DLLA}) and (\ref{ReNDLLA}).
The last term in (\ref{R3OPEContapp}) is currently not available in the BFKL approach and brings in some new information about the next-to-leading corrections to the  BFKL eigenvalue in the adjoint representation.

\setcounter{equation}{0}

\renewcommand{\theequation}{D.\arabic{equation}}

\section{ Contribution of $\gamma_1^{-}(p)$ up to five loops}\label{app:45plus}

In this section we calculate a contribution of  $\gamma^{-}_1(p)$ to the remainder function~(\ref{ROPEexp}) up to five loops. We show that in the Double Leading Logarithmic Approximation~(DLLA)  in the Mandelstam region the main contribution comes from the highest power of $\gamma^{-}_1(p)$ in the integral of (\ref{ROPEexp}), namely $(\gamma^{-}_1(p))^{\ell-1}$, where $\ell$ is a number of loops.  It is useful to define
\beqn\label{gammaminusapp}
\gamma_1(p)=\gamma^{+}_1(p)+\gamma^{-}_1(p), \;\;\;\gamma^{\pm}_1(p)=\psi \left(\frac{3}{2}\pm\frac{ip}{2} \right)-\psi \left(1 \right)
\eeqn
and
\beqn\label{hkMapp}
h^{-}_k (\sigma)=\int_{-\infty}^{\infty} c^{0}(p) \left( \gamma^{-}_1(p) \right)^k e^{ip\sigma} dp.
\eeqn

We use the residue theorem  noting that the function
\beqn\label{fkapp}
f_k(p)=c^{0}(p) \left( \gamma^{-}_1(p) \right)^k e^{ip\sigma}
\eeqn
has second-order poles only  at $p= \pm i$ and all other poles are simple poles. For $\sigma >0$ we close the contour of integration in the upper semiplane,
 where $\gamma^{-}_1(p)$ has no poles. We start with $h^{-}_1 (\sigma)$ relevant for $2$ loops of the AGMSV remainder function in (\ref{ROPEexp}).
  It is easy to find the residue of $f_1 (p)$ at the pole $p=i$
  \beqn\label{Resdouble1}
i2 \pi Res (f_1(p),i)=
4 e^{-\sigma } \sigma +4 e^{-\sigma }-\frac{1}{3} \pi ^2 e^{-\sigma }.
\eeqn
Other poles in the upper semiplane come from  $\cosh \left( \frac{\pi p}{2}\right)$ in $c^{0}(p)$. Using the fact that
all of these poles are simple poles we can make a substitution
\beqn
\frac{1}{\cosh \left( \frac{\pi p}{2}\right)} \rightarrow \sum_{n=1}^{\infty}\frac{i2}{\pi}\frac{(-1)^{n+1} }{p- i(2n+1)},
\eeqn
which  accounts properly for the pole and residue structure of $c^{0}(p)$ for $p \neq  i$ in the upper semiplane.
Then the calculation of the residues of $f_{1}(p)$ at these poles becomes straightforward and we get
\beqn\label{Ressingle1}
&& i2\pi \sum_{n=1}^{\infty} Res(f_1(p),i(2n+1))=-\sum_{n=1}^{\infty}\frac{2 (-1)^n e^{-\sigma -2 n \sigma } (\psi(2+n)-\psi(1))}{n (1+n)}\\
&& =-\sum_{n=1}^{\infty}\frac{2 (-1)^n e^{-\sigma -2 n \sigma } S_{1}(n+1)}{n (1+n)} =
 -4 e^{-\sigma } +4  \cosh \sigma \ln \left(1+e^{-2 \sigma }\right) \nonumber
 \\
&&
 -2 \cosh \sigma \ln ^2\left(1+e^{-2 \sigma }\right)
   -4 \cosh \sigma \text{Li}_2\left(-e^{-2 \sigma }\right), \nonumber
\eeqn
where the we used the identity
\beqn
S_{1}(n)=\psi (n+1)-\psi(1)
\eeqn
for the harmonic number $S_{m}(n)=\sum_{i=1}^{n}1/i^m$ and the polygamma function $\psi(x)=d/dx (\ln  \Gamma(x))$.
 The series summation in  (\ref{Ressingle1}) as well as  other sums in this section is performed using the XSummer package for FORM by Moch and Uwer~\cite{Moch:2001zr,Xsummer}.
Finally, adding (\ref{Ressingle1}) to (\ref{Resdouble1}) we obtain
\beqn\label{h1minusapp}
&&
h^{-}_1 (\sigma)=\int_{-\infty}^{\infty} c^{0}(p)  \gamma^{-}_1(p) e^{ip\sigma} dp=
4 e^{-\sigma } \sigma -\frac{1}{3} \pi ^2 e^{-\sigma }
+4  \cosh \sigma \ln \left(1+e^{-2 \sigma }\right)
 \\
&&
 -2 \cosh \sigma \ln ^2\left(1+e^{-2 \sigma }\right)
   -4 \cosh \sigma \text{Li}_2\left(-e^{-2 \sigma }\right), \nonumber
\eeqn
which after the analytic continuation along the path $\mathbf{B}$ results in
\beqn
\Delta_1^{-}(\sigma) =-4 i \pi  e^{\sigma }+8 i \pi  \sigma  \cosh \sigma +8 i \pi  \cosh \sigma  \ln (2 \cosh \sigma )
\eeqn
for an arbitrary positive $\sigma$, where
\beqn
h^{-}_k (\sigma) \Rightarrow -h^{-}_k (\sigma)+\Delta_k^{-}(\sigma).
\eeqn
In the Regge limit ($\sigma\simeq -1/2 \ln (1-u_1) \to \infty$)  after the analytic continuation we obtain
 \beqn
h^{-}_1 (\sigma) \Longrightarrow -\frac{4 i \pi }{\sqrt{1-u_1}}-\frac{4 i \pi  \ln (1-u_1)}{\sqrt{1-u_1}}
 \eeqn
and the remainder function (\ref{ROPEexp}) at two loops at the double logarithmic accuracy reads
\beqn\label{rd2app}
 &&R^{(2)-}_{OPE} =   -\cos \phi \; e^{-\tau}\;\tau h^{-}_1 (\sigma) \Longrightarrow
-i \pi  \cos  (\phi_2-\phi_3)  |w|\ln  ^2(1-u_1)
\\
&&
-i \pi  \cos  (\phi_2-\phi_3)  |w|\ln  (1-u_1)-2 i \pi  \cos  (\phi_2-\phi_3)  |w|\ln  (1-u_1) \ln  |w| \nonumber
\\
 &&
+2 i \pi   \cos  (\phi_2-\phi_3) |w|\ln  (2) \ln  (1-u_1)-2 i \pi \cos  (\phi_2-\phi_3)
    |w|\ln  |w|+2 i \pi  |w|\ln  2 \nonumber
\\
&&
\simeq
-i2  \pi  \cos  (\phi_2-\phi_3)  |w| \ln |w| \ln (1-u_1)    -i2  \pi \cos  (\phi_2-\phi_3)   |w|\ln |w|. \nonumber
\eeqn
The expression in (\ref{rd2app}) reproduces the BFKL result, despite the fact that we considered only $\gamma^{-}_{1}(p)$
part~(see (\ref{gammaminusapp})) of the anomalous dimension $\gamma_1(p)$ in (\ref{gamma}).

In a similar way we calculate the contribution from $\gamma_1^{-}(p)$ at three loops. First we close the contour in the upper semiplane and find
the residue of $f_2(p)$ in (\ref{fkapp}) at the second-order pole   $p=i$
  \beqn\label{Resdouble2}
i2 \pi Res (f_2(p),i)=
4 e^{-\sigma } \sigma +6 e^{-\sigma }-\frac{2}{3} \pi ^2 e^{-\sigma }.
\eeqn
Next we calculate the residue of $f_2(p)$ at the simple poles in the upper semiplane for $p \neq i$
\beqn\label{Ressingle2}
&& i2\pi \sum_{n=1}^{\infty} Res(f_2(p),i(2n+1))=-\sum_{n=1}^{\infty}\frac{2 (-1)^n e^{-\sigma -2 n \sigma } (\psi(2+n)-\psi(1))^2}{n (1+n)}\\
&& =-\sum_{n=1}^{\infty}\frac{2 (-1)^n e^{-\sigma -2 n \sigma } S^2_{1}(n+1)}{n (1+n)} =
-2 e^{-\sigma } \text{Li}_2\left(-e^{-2 \sigma }\right)-4  \cosh \sigma\text{Li}_2\left(-e^{-2 \sigma }\right) \nonumber
\\
&& -4  \cosh \sigma\text{Li}_3\left(-e^{-2 \sigma }\right)  +4  \cosh \sigma \ln \left(1+e^{-2 \sigma }\right)
   \text{Li}_2\left(-e^{-2 \sigma }\right)    -6 e^{-\sigma }
   \nonumber
   \\
   &&+\frac{4}{3}\cosh \sigma \ln ^3\left(1+e^{-2 \sigma }\right)   -4 \cosh \sigma
   \ln ^2\left(1+e^{-2 \sigma }\right)   +4 \cosh \sigma \ln \left(1+e^{-2 \sigma }\right). \nonumber
\eeqn
The expression in (\ref{Resdouble2}) together with (\ref{Ressingle2}) gives the required integral
\beqn \label{h2minusapp}
&&
h^{-}_2 (\sigma)=\int_{-\infty}^{\infty} c^{0}(p)  (\gamma^{-}_1(p))^2 e^{ip\sigma} dp=
2 e^{\sigma } \text{Li}_2\left(-e^{-2 \sigma }\right)-8 \cosh \sigma \text{Li}_2\left(-e^{-2 \sigma }\right)
\\
&&
-4 \cosh \sigma \text{Li}_3\left(-e^{-2
   \sigma }\right) +4 \cosh \sigma \ln \left(1+e^{-2 \sigma }\right) \text{Li}_2\left(-e^{-2 \sigma }\right)  -4 e^{\sigma }
   \sigma +\frac{2 \pi ^2 e^{\sigma }}{3}
+8 \sigma  \cosh \sigma \nonumber
\\
&&-\frac{4}{3}  \pi ^2 \cosh \sigma  +\frac{4}{3} \cosh \sigma \ln ^3\left(1+e^{-2 \sigma
   }\right) -4 \cosh \sigma \ln ^2\left(1+e^{-2 \sigma }\right)  +4  \cosh \sigma\ln \left(1+e^{-2 \sigma }\right). \nonumber
\eeqn
After the analytic continuation of $h^{-}_2 (\sigma)$ along the path $\mathbf{B}$ we get
\beqn
&& \Delta_2^{-}(\sigma) = -8 i \pi \cosh \sigma \text{Li}_2\left(-e^{-2 \sigma }\right) -8 i \pi  \sigma ^2 \cosh \sigma-8 i \pi  e^{\sigma } \sigma -4 \pi ^2    e^{\sigma }-4 i \pi  e^{\sigma }
\\ &&
+8 \pi ^2 \sigma  \cosh \sigma+16 i \pi  \sigma  \cosh \sigma
-\frac{4}{3} i \pi ^3 \cosh \sigma-8 i
   \pi  \cosh \sigma \ln  ^2(2 \cosh \sigma) \nonumber
\\ &&
+8 \pi ^2 \cosh \sigma \ln  (2 \cosh \sigma)+16 i \pi  \cosh \sigma \ln  (2 \cosh
   \sigma). \nonumber
\eeqn
In the Regge limit ($\sigma\simeq -1/2 \ln (1-u_1) \to \infty$)  after the analytic continuation we obtain
 \beqn
&&h^{-}_2(\sigma) \Longrightarrow
-\frac{2 i \pi ^3}{3 \sqrt{1-u_1}}-\frac{4 \pi ^2}{\sqrt{1-u_1}}-\frac{4 i \pi }{\sqrt{1-u_1}}-\frac{2 i \pi  \ln^2(1-u_1)}{\sqrt{1-u_1}}
\\
&&-\frac{4
   \pi ^2 \ln(1-u_1)}{\sqrt{1-u_1}}-\frac{4 i \pi  \ln(1-u_1)}{\sqrt{1-u_1}} \nonumber
\eeqn

and the remainder function (\ref{ROPEexp}) at three loops in the double logarithmic approximation reads
\beqn\label{rd3app}
 &&R^{(3)-}_{OPE} =   \cos \phi \; e^{-\tau}\;\frac{\tau^2}{2} h^{-}_2 (\sigma) \Longrightarrow
-\frac{i \pi}{2}  \cos  (\phi_2-\phi_3)   |w| \ln ^2|w| \ln ^2(1-u_1)
\\
&&-\pi ^2  \cos  (\phi_2-\phi_3)  |w| \ln ^2|w| \ln (1-u_1)  -i \pi  \cos  (\phi_2-\phi_3)  |w| \ln (1-u_1). \nonumber
\eeqn
For simplicity of the presentation we retain only the leading and the next-to-leading terms in $\ln |w|$ in (\ref{rd3app}).  The first two terms in RHS of (\ref{rd3app}) reproduce the BFKL result in (\ref{DLLA}) and (\ref{ReNDLLA}), while the last term is not captured by the LLA BFKL analysis.

Using the same procedure we calculate   the contribution of $\gamma_1^{-}(p)$ at four loops. First we find the residue of $f_3(p)$ at the  $p=i$
\beqn\label{Resdouble3}
i2 \pi Res (f_3(p),i)=e^{-\sigma } \left(8-\pi ^2+4 \sigma \right)
\eeqn
and then at other (simple) poles in the upper semiplane
\beqn\label{Ressingle3}
&& i2\pi \sum_{n=1}^{\infty} Res(f_3(p),i(2n+1))=-\sum_{n=1}^{\infty}\frac{2 (-1)^n e^{-\sigma -2 n \sigma } (\psi(2+n)-\psi(1))^3}{n (1+n)}\\
&& =-\sum_{n=1}^{\infty}\frac{2 (-1)^n e^{-\sigma -2 n \sigma } S^3_{1}(n+1)}{n (1+n)}=
-8 e^{-\sigma }+\frac{4}{15} \pi ^4 \cosh \sigma-e^{\sigma } \pi ^2 \ln \left(1+e^{-2\sigma}\right)\nonumber
\\
&&+4 \cosh \sigma \ln \left(1+e^{-2\sigma}\right)
 -6 e^{-\sigma } \sigma  \ln^2 \left(1+e^{-2\sigma}\right)
-6 \cosh \sigma \ln^2 \left(1+e^{-2\sigma}\right)+12 \sigma  \cosh \sigma \ln^2 \left(1+e^{-2\sigma}\right) \nonumber
\\
&&+2 e^{-\sigma } \ln^3 \left(1+e^{-2\sigma}\right)+3 e^{\sigma } \ln^3 \left(1+e^{-2\sigma}\right) \nonumber
+4 \sigma  \cosh \sigma \ln^3 \left(1+e^{-2\sigma}\right)
-6 e^{-\sigma } \text{Li}_2\left(-e^{-2 \sigma}\right)
\\
&&-2 e^{\sigma } \text{Li}_2\left(-e^{-2 \sigma}\right)+6 e^{-\sigma } \ln \left(1+e^{-2\sigma}\right) \text{Li}_2\left(-e^{-2 \sigma}\right) \nonumber
-6 \cosh \sigma \ln^2 \left(1+e^{-2\sigma}\right) \text{Li}_2\left(-e^{-2 \sigma}\right)
\\
&&-4 \cosh \sigma \text{Li}_2\left(-e^{-2 \sigma}\right)^2-6 e^{-\sigma } \text{Li}_3\left(-e^{-2 \sigma}\right)-2 e^{\sigma } \text{Li}_3\left(-e^{-2 \sigma}\right)\nonumber
+4 \cosh \sigma \ln \left(1+e^{-2\sigma}\right) \text{Li}_3\left(-e^{-2 \sigma}\right)
\\
&&-6 e^{\sigma } \text{Li}_3\left(\frac{1}{1+e^{-2 \sigma}}\right)-12 \cosh \sigma \ln \left(1+e^{-2\sigma}\right) \text{Li}_3\left(\frac{1}{1+e^{-2 \sigma}}\right) \nonumber
-4 \cosh \sigma \text{Li}_4\left(-e^{-2 \sigma}\right)
\\
&&-24 \cosh \sigma \text{Li}_4\left(\frac{1}{1+e^{-2 \sigma}}\right)+12 \cosh \sigma \text{Li}_{2,2}\left(-e^{-2 \sigma}\right)+6 e^{\sigma } \zeta_3 \nonumber
-12 \cosh \sigma \ln \left(1+e^{-2\sigma}\right) \zeta_3. \nonumber
\eeqn

Adding the contributions of the second-order pole in (\ref{Resdouble3}) and simple poles in (\ref{Ressingle3}) we readily obtain
\beqn\label{h3minusapp}
&&
h^{-}_3 (\sigma)=\int_{-\infty}^{\infty} c^{0}(p) \left( \gamma^{-}_1(p) \right)^3 e^{ip\sigma} dp=
-\pi^2e^{-\sigma } +4 \sigma e^{-\sigma }+\frac{4}{15} \pi ^4 \cosh \sigma-e^{\sigma } \pi ^2 \ln \left(1+e^{-2\sigma}\right)
\\
&&+4 \cosh \sigma \ln \left(1+e^{-2\sigma}\right)
 -6 e^{-\sigma } \sigma  \ln^2 \left(1+e^{-2\sigma}\right)
-6 \cosh \sigma \ln^2 \left(1+e^{-2\sigma}\right)+12 \sigma  \cosh \sigma \ln^2 \left(1+e^{-2\sigma}\right) \nonumber
\\
&&+2 e^{-\sigma } \ln^3 \left(1+e^{-2\sigma}\right)+3 e^{\sigma } \ln^3 \left(1+e^{-2\sigma}\right) \nonumber
+4 \sigma  \cosh \sigma \ln^3 \left(1+e^{-2\sigma}\right)
-6 e^{-\sigma } \text{Li}_2\left(-e^{-2 \sigma}\right)
\\
&&-2 e^{\sigma } \text{Li}_2\left(-e^{-2 \sigma}\right)+6 e^{-\sigma } \ln \left(1+e^{-2\sigma}\right) \text{Li}_2\left(-e^{-2 \sigma}\right) \nonumber
-6 \cosh \sigma \ln^2 \left(1+e^{-2\sigma}\right) \text{Li}_2\left(-e^{-2 \sigma}\right)
\\
&&-4 \cosh \sigma \text{Li}_2\left(-e^{-2 \sigma}\right)^2-6 e^{-\sigma } \text{Li}_3\left(-e^{-2 \sigma}\right)-2 e^{\sigma } \text{Li}_3\left(-e^{-2 \sigma}\right)\nonumber
+4 \cosh \sigma \ln \left(1+e^{-2\sigma}\right) \text{Li}_3\left(-e^{-2 \sigma}\right)
\\
&&-6 e^{\sigma } \text{Li}_3\left(\frac{1}{1+e^{-2 \sigma}}\right)-12 \cosh \sigma \ln \left(1+e^{-2\sigma}\right) \text{Li}_3\left(\frac{1}{1+e^{-2 \sigma}}\right) \nonumber
-4 \cosh \sigma \text{Li}_4\left(-e^{-2 \sigma}\right)
\\
&&-24 \cosh \sigma \text{Li}_4\left(\frac{1}{1+e^{-2 \sigma}}\right)+12 \cosh \sigma \text{Li}_{2,2}\left(-e^{-2 \sigma}\right)+6 e^{\sigma } \zeta_3 \nonumber
-12 \cosh \sigma \ln \left(1+e^{-2\sigma}\right) \zeta_3. \nonumber
\eeqn

In a similar way we calculate also $h^{-}_4 (\sigma)$ needed for the five-loop remainder function in (\ref{ROPEexp}).
First we find a contribution from the second-order pole of $f_4 (p)$
\beqn\label{Resdouble4}
i2 \pi Res (f_4(p),i)=\frac{2}{3} e^{-\sigma } \left(15-2 \pi ^2+6 \sigma \right)
\eeqn
and then simples poles
\beqn\label{Ressingle4}
&& i2\pi \sum_{n=1}^{\infty} Res(f_4(p),i(2n+1))=-\sum_{n=1}^{\infty}\frac{2 (-1)^n e^{-\sigma -2 n \sigma } (\psi(2+n)-\psi(1))^4}{n (1+n)}\\
&& =-\sum_{n=1}^{\infty}\frac{2 (-1)^n e^{-\sigma -2 n \sigma } S^4_{1}(n+1)}{n (1+n)}. \nonumber
\eeqn

The series in (\ref{Ressingle4}) is also summed using  the XSummer package for  FORM and together with the double-pole contribution in (\ref{Resdouble4}) it
results in

\beqn\label{h4minusapp}
&& h^{-}_4 (\sigma)=\int_{-\infty}^{\infty} c^{0}(p) \left( \gamma^{-}_1(p) \right)^4 e^{ip\sigma} dp =\frac{4}{15} e^{-\sigma } \pi ^4
-\frac{4}{3} e^{-\sigma } \pi ^2+16 e^{\sigma } \zeta_3\\
&&-\frac{6}{5} \cosh \sigma  \ln  ^5\left(1+e^{-2 \sigma }\right)-3 e^{-\sigma } \ln  ^4\left(1+e^{-2 \sigma }\right)-8 \sigma  \cosh \sigma
   \ln  ^4\left(1+e^{-2 \sigma }\right)\nonumber
\\ \nonumber
&&
+6 \cosh \sigma  \ln  ^4\left(1+e^{-2 \sigma }\right)-\frac{8}{3} e^{-\sigma } \ln  ^3\left(1+e^{-2
   \sigma }\right)-8 e^{-\sigma } \sigma  \ln  ^3\left(1+e^{-2 \sigma }\right)
\\ \nonumber
&&
+32 \sigma  \cosh \sigma  \ln  ^3\left(1+e^{-2 \sigma
   }\right)
+\frac{40}{3} \cosh \sigma  \ln  ^3\left(1+e^{-2 \sigma }\right)+8 \cosh \sigma  \text{Li}_2\left(-e^{-2 \sigma }\right) \ln
   ^3\left(1+e^{-2 \sigma }\right)
\\ \nonumber
&&
+16 e^{\sigma } \sigma  \ln  ^2\left(1+e^{-2 \sigma }\right)-8 \cosh \sigma  \ln  ^2\left(1+e^{-2 \sigma
   }\right)-24 \cosh \sigma  \text{Li}_2\left(-e^{-2 \sigma }\right) \ln  ^2\left(1+e^{-2 \sigma }\right)
\\ \nonumber
&&
+12 e^{\sigma }
   \text{Li}_2\left(\frac{1}{1+e^{-2 \sigma }}\right) \ln  ^2\left(1+e^{-2 \sigma }\right)-8 \cosh \sigma  \text{Li}_3\left(-e^{-2 \sigma
   }\right) \ln  ^2\left(1+e^{-2 \sigma }\right)
\\ \nonumber
&&+24 \cosh \sigma  \text{Li}_3\left(\frac{1}{1+e^{-2 \sigma }}\right) \ln  ^2\left(1+e^{-2
   \sigma }\right)
+24 \cosh \sigma  \zeta_3 \ln  ^2\left(1+e^{-2 \sigma }\right)
\\ \nonumber
&&+4 \cosh \sigma  \text{Li}_2\left(-e^{-2 \sigma
   }\right){}^2 \ln  \left(1+e^{-2 \sigma }\right)-\frac{8}{15} \pi ^4 \cosh \sigma  \ln  \left(1+e^{-2 \sigma }\right)-\frac{16}{3} \pi ^2
   \cosh \sigma  \ln  \left(1+e^{-2 \sigma }\right)
\\ \nonumber
&&
+4 \cosh \sigma  \ln  \left(1+e^{-2 \sigma }\right)+16 e^{-\sigma }
   \text{Li}_2\left(-e^{-2 \sigma }\right) \ln  \left(1+e^{-2 \sigma }\right)-8 \cosh \sigma  \text{Li}_2\left(-e^{-2 \sigma }\right) \ln
   \left(1+e^{-2 \sigma }\right)
\\ \nonumber
&&
+16 \cosh \sigma  \text{Li}_3\left(-e^{-2 \sigma }\right) \ln  \left(1+e^{-2 \sigma }\right)-24 e^{-\sigma }
   \text{Li}_3\left(\frac{1}{1+e^{-2 \sigma }}\right) \ln  \left(1+e^{-2 \sigma }\right)
\\ \nonumber
&&
+4 \cosh \sigma  \text{Li}_4\left(-e^{-2 \sigma
   }\right) \ln  \left(1+e^{-2 \sigma }\right)+48 \cosh \sigma  \text{Li}_4\left(\frac{1}{1+e^{-2 \sigma }}\right) \ln  \left(1+e^{-2 \sigma
   }\right)
\\ \nonumber
&&
-32 \cosh \sigma   \text{Li}_{2,2}\left(-e^{-2 \sigma }\right) \ln  \left(1+e^{-2 \sigma }\right)-48 \cosh \sigma  \zeta_3 \ln
   \left(1+e^{-2 \sigma }\right)+\frac{8}{3} e^{-\sigma } \pi ^2 \ln  \left(1+e^{-2 \sigma }\right)
\\ \nonumber
&&-6 e^{-\sigma } \text{Li}_2\left(-e^{-2
   \sigma }\right){}^2-4 \cosh \sigma  \text{Li}_2\left(-e^{-2 \sigma }\right){}^2+4 e^{-\sigma } \sigma +\frac{8}{15} \pi ^4 \cosh (\sigma
   )
\\ \nonumber
&&
-8 \cosh \sigma
H_{0,0,0,1,1}\left(-e^{-2 \sigma }\right)-20 \cosh \sigma
H_{0,0,1,0,1}\left(-e^{-2 \sigma }\right)
-48 \cosh \sigma
  H_{0,0,0,1,1}\left(-e^{-2 \sigma }\right)
\\ \nonumber
&&
-12 \cosh \sigma
H_{0,1,0,0,1}\left(-e^{-2 \sigma }\right)
-16 \cosh \sigma
H_{0,1,0,1,1}\left(-e^{-2 \sigma }\right)
-16 \cosh \sigma  H_{0,1,1,0,1}\left(-e^{-2 \sigma }\right)
\\ \nonumber
&&
-6 e^{-\sigma } \text{Li}_2\left(-e^{-2 \sigma }\right)-4 \cosh
   (\sigma ) \text{Li}_2\left(-e^{-2 \sigma }\right)
-10 e^{-\sigma } \text{Li}_3\left(-e^{-2 \sigma }\right)
\\ \nonumber \nonumber
&&
-4 \cosh \sigma
   \text{Li}_3\left(-e^{-2 \sigma }\right)-16 e^{\sigma } \text{Li}_3\left(\frac{1}{1+e^{-2 \sigma }}\right)
-6 e^{-\sigma }
   \text{Li}_4\left(-e^{-2 \sigma }\right)
\\ \nonumber
&&
-4 \cosh \sigma  \text{Li}_4\left(-e^{-2 \sigma }\right)+24 e^{\sigma }
   \text{Li}_4\left(\frac{1}{1+e^{-2 \sigma }}\right)-96 \cosh \sigma  \text{Li}_4\left(\frac{1}{1+e^{-2 \sigma }}\right)
\\ \nonumber
&&
-4 \cosh \sigma
   \text{Li}_5\left(-e^{-2 \sigma }\right)+8 e^{-\sigma } \text{Li}_{2,2}\left(-e^{-2 \sigma }\right)+32 \cosh \sigma  \text{Li}_{2,2}\left(-e^{-2 \sigma
   }\right).
\eeqn

 The function  $h^{-}_4 (\sigma)$ in (\ref{h4minusapp}) is expressed in terms of the classical polylogarithms, and the harmonic polylogarithms~(HPL)~\cite{Moch:2001zr} $H_{a_1,a_2,...,a_n}(x)$ are recursively defined  by
\beqn\label{HPLapp}
H_{a_1,a_2,...,a_n}(x)=\int_0^x dt g_{a_1}(t) H_{a_2,...,a_n}(t),\;\; a_i=0, \pm 1,
\eeqn
where
\beqn
g_{\pm}(x) =\frac{1}{1\mp x}, \;\; g_0(x)=\frac{1}{x},\;\; H_{\pm}(x)=\mp \ln (1 \mp x), \;\;H_{0}(x)=\ln x
\eeqn
and at least one of the indices $a_i$ is not zero. For all $a_i=0$, one has
\beqn
H_{0,0,...,0}(x)=\frac{1}{n!} \ln^n x.
\eeqn
For a nice introduction and the list of HPL with the transcendentality (number of indices $a_i$) up to four the reader is referred to the appendix of \cite{Smirnov:2004ym}. In particular, we used here
\beqn
H_{0,0,1,1}(x)=\text{Li}_{2,2}(x).
\eeqn

The analytic continuation is similar to that done for two and three loops though due to a complexity of the calculations it is much easier  to perform the continuation together with subsequent Regge limit $\sigma \to + \infty$. In this case only those HPL, which have all rightmost indices $1$ contribute in (\ref{h4minusapp}). Other HPL, where the index $0$ stand to right of the index $1$ all vanish for $\sigma \to +\infty$ after the analytic continuation. As simple example, we analytically continue $H_{1,0,1}\left(-e^{-2\sigma} \right)$, noting that
\beqn \label{expsapp}
-e^{-2\sigma} \simeq -\frac{1-u_1}{u_1}
\eeqn
for the analytic continuation along the path $\mathbf{B}$, where $\sigma \simeq 1/2 \ln (u_1/(1-u_1))$. In the multi-Regge kinematics $|u_1| \to 1^{-}$ and the expression in (\ref{expsapp}) rotates in the anti-clockwise direction around the origin  as explained in the previous section. Using the integral representation of HPL (\ref{HPLapp}) we perform the analytic continuation
\beqn
&&H_{1,0,1}\left(-e^{-2\sigma} \right)=\int_0^{-\frac{1-u_1}{u_1}} \frac{dt}{1-t} \int_0^{t} \frac{dt'}{t'}\int_0^{t'} \frac{dt''}{1-t''}\Rightarrow
\\
&&
\int_0^{-\frac{1-|u_1|}{|u_1|}} \frac{dt}{1-t} \int_1^{t} \frac{dt'}{t'}-i2\pi  \int_0^{-\frac{1-|u_1|}{|u_1|}} \frac{dt'}{t'}\int_0^{t'} \frac{dt''}{1-t''}-i2\pi \int_0^{-\frac{1-|u_1|}{|u_1|}}\frac{dt}{1-t}\int_1^t\frac{dt'}{t'} \nonumber\\
&& =H_{1,0,1}\left(-e^{-2\sigma} \right) -i2 \pi H_{0,1}\left(-e^{-2\sigma} \right)-i2\pi H_{1,0}\left(-e^{-2\sigma} \right)  \overbrace{\longrightarrow}^{\text{Regge limit}} 0.
\eeqn
Note that the singularities of HPL functions are determined by the rightmost index. All HPL appearing in (\ref{h4minusapp}) have the same rightmost index $1$. This means that the branch cut of all HPL appearing in (\ref{h4minusapp}) is the same as the branch cut of $H_{1}(-e^{-2 \sigma})=-\ln\left(1+e^{-2 \sigma} \right)$.  The classical polylogarithms  $\text{Li}_{n} \left(-e^{-2 \sigma} \right)$ have the same cut structure as well, because they are a special case of HPL with the rightmost index $1$, namely
\beqn
\text{Li}_{n} \left(-e^{-2 \sigma} \right)=H_{0,0,...,0,1}\left(-e^{-2 \sigma} \right),
\eeqn
where $n$ is the number of indices. For $\text{Li}_{n} \left(-e^{-2 \sigma} \right)$ all but the rightmost indices are $0$, so that they do make a contribution after the analytic continuation in the Regge limit  and  can be compactly written as
\beqn
\text{Li}_{n} \left(-e^{-2 \sigma} \right)  \Rightarrow \text{Li}_{n} \left(-e^{-2 \sigma}\right)-i2\pi \frac{(-2\sigma +i\pi )^{n-1}}{(n-1)!}
\eeqn
for arbitrary value of $\sigma >0$.
In the expressions for $h_3(\sigma)$ and  $h_4(\sigma)$ for the first time we have the generalized Nielsen polylogarithm $\text{Li}_{2,2}\left(-e^{-2 \sigma} \right)$. We write the integral representation of  the corresponding HPL and analytically continue it~(along the path $\mathbf{B}$)  to our region
\beqn
&&\text{Li}_{2,2}\left(-e^{-2 \sigma} \right)=H_{0,0,1,1}\left(-e^{-2 \sigma} \right)
=\int_0^{-\frac{1-u_1}{u_1}} \frac{dt}{t}\int_0^{t} \frac{dt'}{t'} \int_0^{t'} \frac{dt''}{1-t''}\int_0^{t''} \frac{dt'''}{1-t'''}\\
&& \int_0^{-\frac{1-u_1}{u_1}} \frac{dt}{t}\int_0^{t} \frac{dt'}{t'} \frac{1}{2} \ln^2 (1-t') \Rightarrow
\int_0^{-\frac{1-|u_1|}{|u_1|}} \frac{dt}{t}\int_0^{t} \frac{dt'}{t'} \frac{1}{2} \left( \ln (1-t')+i2\pi \right)^2 \nonumber\\
&& =H_{0,0,1,1}\left(-e^{-2 \sigma} \right)-i2\pi H_{0,0,1}\left(-e^{-2 \sigma} \right)+\frac{(i2\pi)^2}{2}H_{0,0}\left(-e^{-2 \sigma} \right)
\overbrace{\longrightarrow}^{\text{Regge limit}} \frac{(i2\pi)^2}{2}\frac{1}{2} \left(-2\sigma+i\pi\right)^2.\nonumber
\eeqn

There are also classical polylogarithms of the argument
\beqn
\frac{1}{1+e^{-2\sigma}} \simeq u_1
\eeqn
in  $h_3(\sigma)$ and  $h_4(\sigma)$. However, they remain the same after the analytic continuation because the circular  path of the continuation $u_1=|u_1| e^{i\psi}$ with $0 \leq \psi \leq -i2\pi $ never crosses the singularities of these functions. Using these  results we readily find the contribution of the maximal powers of $\gamma_1^{-}(p)$ in the integrand of  the remainder function~(\ref{ROPEexp}). In the Mandestam region in the multi-Regge kinematics we get
\beqn
&& h^{-}_3 (\sigma) \Longrightarrow  \frac{24 i \pi  \zeta_3}{\sqrt{1-u_1}}-\frac{8 \pi ^4}{3 \sqrt{1-u_1}}+\frac{4 i \pi ^3}{\sqrt{1-u_1}}-\frac{8 \pi
   ^2}{\sqrt{1-u_1}}-\frac{4 i \pi }{\sqrt{1-u_1}}-\frac{2 i \pi  \ln^3(1-u_1)}{3 \sqrt{1-u_1}} \hspace{1.9cm}
\\
&&-\frac{4 \pi ^2 \ln
   ^2(1-u_1)}{\sqrt{1-u_1}}-\frac{2 i \pi  \ln^2(1-u_1)}{\sqrt{1-u_1}}+\frac{6 i \pi ^3 \ln
   (1-u_1)}{\sqrt{1-u_1}}-\frac{8 \pi ^2 \ln(1-u_1)}{\sqrt{1-u_1}}-\frac{4 i \pi  \ln
   (1-u_1)}{\sqrt{1-u_1}}, \nonumber
\eeqn
which gives the four loop expression
\beqn\label{R4Mapp}
&& R^{(4)-}_{OPE} =   -\cos \phi \; e^{-\tau}\;\frac{\tau^3}{3!} h^{-}_3 (\sigma) \Longrightarrow
  -\frac{i \pi}{18} \cos (\phi_2-\phi_3)   |w| \ln ^3|w| \ln ^3(1-u_1)  \;\;\;\;\;\;\;\;\;\;
\\
&&-\frac{\pi ^2}{3} \cos (\phi_2-\phi_3)  |w|\ln ^3 |w| \ln ^2(1-u_1) -\frac{i\pi}{6} \cos (\phi_2-\phi_3)  |w| \ln ^3 |w|\ln
   ^2(1-u_1).   \nonumber
\eeqn
In (\ref{R4Mapp}) we retain only the leading and next-to-leading   powers of $\ln (1-u_1)$ in terms leading in $\ln |w|$.
At five loops for $\sigma \to \infty$ we obtain
\beqn
&& h^{-}_4 (\sigma) \Longrightarrow \frac{48 \pi ^2 \zeta_3}{\sqrt{1-u_1}}+\frac{48 i \pi  \zeta_3}{\sqrt{1-u_1}}-\frac{16 \zeta_3}{\sqrt{1-u_1}}-\frac{353
   i \pi ^5}{6 \sqrt{1-u_1}}+\frac{678 \pi ^4}{5 \sqrt{1-u_1}}-\frac{154 i \pi ^3}{3 \sqrt{1-u_1}}
\\
&&
-\frac{12 \pi
   ^2}{\sqrt{1-u_1}}
-\frac{4 i \pi }{\sqrt{1-u_1}}-\frac{i \pi  \ln  ^4(1-u_1)}{6 \sqrt{1-u_1}}-\frac{2 \pi ^2 \ln
   ^3(1-u_1)}{\sqrt{1-u_1}}-\frac{2 i \pi  \ln  ^3(1-u_1)}{3 \sqrt{1-u_1}}+\frac{9 i \pi ^3 \ln
   ^2(1-u_1)}{\sqrt{1-u_1}} \nonumber
\\
&&
-\frac{6 \pi ^2 \ln  ^2(1-u_1)}{\sqrt{1-u_1}}-\frac{2 i \pi  \ln
   ^2(1-u_1)}{\sqrt{1-u_1}}+\frac{22 \pi ^4 \ln  (1-u_1)}{3 \sqrt{1-u_1}}+\frac{18 i \pi ^3 \ln
   (1-u_1)}{\sqrt{1-u_1}}-\frac{12 \pi ^2 \ln  (1-u_1)}{\sqrt{1-u_1}} \nonumber
\\
&&
-\frac{4 i \pi  \ln
   (1-u_1)}{\sqrt{1-u_1}} \simeq -\frac{i \pi  \ln ^4(1-u_1)}{6 \sqrt{1-u_1}}-\frac{2 \pi ^2 \ln ^3(1-u_1)}{\sqrt{1-u_1}}-\frac{2 i \pi  \ln
   ^3(1-u_1)}{3 \sqrt{1-u_1}},
\nonumber
\eeqn
which gives
\beqn\label{R5Mapp}
&& R^{(5)-}_{OPE} =   \cos \phi \; e^{-\tau}\;\frac{\tau^4}{4!} h^{-}_4 (\sigma) \Longrightarrow
-\frac{i \pi}{288} \cos  (\phi_2-\phi_3)    |w| \ln^4 |w| \ln^4(1-u_1)
\\
&& -\frac{ \pi ^2}{24}\cos  (\phi_2-\phi_3)  |w|  \ln^4 |w| \ln^3(1-u_1)
-\frac{i\pi}{72}  \cos  (\phi_2-\phi_3)  |w| \ln^4 |w| \ln^3(1-u_1).
\nonumber
\eeqn
As in the previous case we leave only the leading and the next-to-leading powers of $\ln(1-u_1)$ in terms leading in $\ln |w|$.
The first terms  in RHS of both  (\ref{R4Mapp}) and (\ref{R5Mapp}) reproduce the BFKL result  in the Double Leading Logarithmic Approximation given by (\ref{DLLA}) and  (\ref{ReNDLLA}).
The last term in RHS of both (\ref{R4Mapp}) and (\ref{R5Mapp}) is currently not available in the LLA BFKL approach and requires a knowledge of the next-to-leading BFKL intercept in the adjoint representation.
We showed that the BFKL prediction in the Double Leading Logarithmic Approximation  can  be reproduced up to five loops by analytically continuing the OPE expression for the remainder function~(\ref{ROPEexp}) to the Mandelstam region,  taking into account only the maximal powers of $\gamma_1^{-}(p)$.

On the other hand any power of $\gamma_{1}^{+}(p)$ in the integrand of (\ref{ROPEexp}) makes the corresponding part of the remainder function to be suppressed by one power of $\ln (1-u_1)$ in the Mandelstam region. These subleading contribution are not captured by the BFKL analysis presented  here and require a knowledge of the next-to-leading BFKL intercept in the adjoint representation as it has been mentioned before.  As a simple example of the above statement, we calculate   the contribution of $\gamma_{1}^{+}(p)$ at two and three loops.  For this purpose we generalize the definition of $h_{k} (\sigma)$ in (\ref{hkMapp}) to include also powers of  $\gamma_{1}^{+}(p)$ in the integrand
\beqn\label{hkMPapp}
h^{-,..,+}_k (\sigma)=\int_{-\infty}^{\infty} c^{0}(p) \left( \gamma^{-}_1(p) \right)^m \left( \gamma^{+}_1(p) \right)^{k-m} e^{ip\sigma} dp.
\eeqn
From the corresponding expression for $\gamma_1(p)=\gamma^{+}_1(p)+\gamma^{-}_1(p)$ given by $h_1 (\sigma)$ in (\ref{h11}) and  $h^{-}_1 (\sigma)$ in (\ref{h1minusapp}) we readily obtain
\beqn\label{h1kPapp}
&& h^{+}_1 (\sigma)=h_1 (\sigma)-h^{-}_1 (\sigma)=4 \cosh  \sigma \text{Li}_2\left(-e^{-2 \sigma }\right)  +6 \sigma ^2 \cosh  \sigma +4 e^{-\sigma } \sigma +\frac{1}{3} \pi ^2 e^{-\sigma }
\\
&&-4
   \sigma  \cosh  \sigma -2 \cosh  \sigma  \ln  ^2(2 \cosh  \sigma )-4 \sigma  \cosh  \sigma  \ln  (2 \cosh  \sigma )+4 \cosh  \sigma
   \ln  (2 \cosh  \sigma ), \nonumber
\eeqn
which after the analytic continuation along the path $\mathbf{B}$ gives
\beqn
h_1^{+} (\sigma) \Rightarrow -h_1^{+} (\sigma)  +\Delta_{1}^{+} (\sigma), \;\; \Delta_{1}^{+} (\sigma)=-4 i \pi  e^{\sigma }.
\eeqn
The leading  contribution  of the remainder function (\ref{ROPEexp}) in the Mandelstam channel then is given by
\beqn
&&R^{(2)+}_{OPE} =-\cos \phi e^{-\tau} \tau h^{+}_{1} (\sigma) \Longrightarrow  -i \pi   \cos  (\phi_2-\phi_3)  |w| \ln  (1-u_1)
\\
&&-i 2 \pi  \cos  (\phi_2-\phi_3)   |w| \ln  |w|
  +i2 \pi  \cos  (\phi_2-\phi_3)   |w| \ln  2, \nonumber
\eeqn
which is suppressed by at least one power of $\ln(1-u_1)$ with respect to the leading term in $R^{(2)-}_{OPE} $ of  (\ref{rd2app}).

This suppression holds also at three loops as described below.  Using  $h^{-}_2 (\sigma)=h^{--}_2 (\sigma)$ in (\ref{h2minusapp}) we can easily obtain
\beqn \label{h2plusappdef}
h^{++}_2 (\sigma)=\int_{-\infty}^{\infty} c^{0}(p)  (\gamma^{+}_1(p))^2 e^{ip\sigma} dp
\eeqn
by complex conjugation of $h^{-}_2 (\sigma)$ with subsequent substitution $\sigma \to -\sigma$
\beqn\label{h2plusapp}
&&h^{++}_2 (\sigma)=h^{--}_2 (-\sigma)=
-2 e^{-\sigma } \text{Li}_2\left(-e^{-2 \sigma }\right)+8 \text{Li}_2\left(-e^{-2 \sigma }\right) \cosh  \sigma -4 \text{Li}_3\left(-e^{-2
   \sigma }\right) \cosh  \sigma
\\
&& -4 \text{Li}_2\left(-e^{-2 \sigma }\right) \ln  \left(1+e^{2 \sigma }\right) \cosh  \sigma +\frac{16}{3}
   \sigma ^3 \cosh  \sigma -4 e^{-\sigma } \sigma ^2
+16 \sigma ^2 \cosh  \sigma \nonumber
\\
&& -8 \sigma ^2 \ln  \left(1+e^{2 \sigma }\right) \cosh (\sigma
   )+4 e^{-\sigma } \sigma +\frac{1}{3} \pi ^2 e^{-\sigma }+\frac{4}{3} \pi ^2 \sigma  \cosh  \sigma
-8 \sigma  \cosh  \sigma \nonumber
\\
&&+\frac{4}{3}
   \ln  ^3\left(1+e^{2 \sigma }\right) \cosh  \sigma -4 \ln  ^2\left(1+e^{2 \sigma }\right) \cosh  \sigma -\frac{2}{3} \pi ^2 \ln  \left(1+e^{2
   \sigma }\right) \cosh  \sigma +4 \ln  \left(1+e^{2 \sigma }\right) \cosh  \sigma.  \nonumber
\eeqn

The mixed term $h^{+-}_2 (\sigma)$  defined by
\beqn \label{h2plusminusapp}
h^{+-}_2 (\sigma)=\int_{-\infty}^{\infty} c^{0}(p)   \gamma^{+}_1(p)  \gamma^{-}_1(p) e^{ip\sigma} dp
\eeqn
can be readily obtained from $h_2(\sigma)$ in (\ref{h2}),  $h^{-}_2 (\sigma)=h^{--}_2 (\sigma)$ in (\ref{h2minusapp})   and $h^{++}_2 (\sigma)$ in (\ref{h2plusapp})
\beqn
&& h^{+-}_2 (\sigma)=\frac{1}{2} \left( h_2 (\sigma)-h^{++}_2 (\sigma)-h^{--}_2 (\sigma)\right)=
8 \sigma  \text{Li}_2\left(-e^{-2 \sigma }\right) \cosh  \sigma
\\
&&
+\frac{16}{3} \sigma
   ^3 \cosh  \sigma
 +8 e^{-\sigma } \sigma -\frac{4}{3} \pi ^2 \sigma  \cosh  \sigma -16 \sigma  \cosh  \sigma
+\frac{4}{3} \ln  ^3\left(1+e^{2   \sigma }\right) \cosh  \sigma  \nonumber
\\
&&
-4 \sigma  \ln  ^2\left(1+e^{2 \sigma }\right) \cosh  \sigma
-4 \ln  ^2\left(1+e^{2 \sigma }\right) \cosh
    \sigma +8 \sigma  \ln  \left(1+e^{2 \sigma }\right) \cosh  \sigma  \nonumber
\\
&&
+\frac{2}{3} \pi ^2 \ln  \left(1+e^{2 \sigma }\right) \cosh  \sigma +8
   \ln  \left(1+e^{2 \sigma }\right) \cosh  \sigma  +8 \text{Li}_3\left(-e^{-2 \sigma }\right) \cosh  \sigma . \nonumber
\eeqn
 Next  we perform the analytic continuation of  $h^{+-}_2 (\sigma)$ and $h^{++}_2 (\sigma)$ along path $\mathbf{B}$  obtaining
\beqn
h^{++}_2 (\sigma) \Rightarrow -h^{++}_2 (\sigma)+\Delta^{++}_2 (\sigma), \;\;\Delta^{++}_2 (\sigma)=-4 i \pi  e^{\sigma }
\eeqn
and
\beqn
h^{+-}_2 (\sigma) \Rightarrow -h^{+-}_2 (\sigma)+\Delta^{+-}_2 (\sigma),
\eeqn
where
\beqn
&& \Delta^{+-}_2 (\sigma)=8 i \pi  \text{Li}_2\left(-e^{-2 \sigma }\right) \cosh  \sigma +16 i \pi  \sigma ^2 \cosh  \sigma +8 i \pi  e^{-\sigma }+\frac{4}{3} i \pi ^3
   \cosh  \sigma
\\
&& -16 i \pi  \cosh  \sigma -4 i \pi  \ln  ^2\left(1+e^{2 \sigma }\right) \cosh  \sigma +8 i \pi  \ln  \left(1+e^{2 \sigma
   }\right) \cosh  \sigma.  \nonumber
\eeqn
In the Regge limit $\sigma \to \infty$ this reads
\beqn\label{h2pppm}
h^{++}_2 (\sigma)  \Longrightarrow -\frac{4 i \pi }{\sqrt{1-u_1}}, \;\;
h^{+-}_2 (\sigma)  \Longrightarrow
\frac{2 i \pi ^3}{3 \sqrt{1-u_1}}-\frac{8 i \pi }{\sqrt{1-u_1}}-\frac{4 i \pi  \ln (1-u_1)}{\sqrt{1-u_1}}.
\eeqn
Finally plugging (\ref{h2pppm}) in $R_{OPE}^{(3)++}$ and $R_{OPE}^{(3)+-}$  defined by
\beqn
R_{OPE}^{(3)++}=\cos \phi \;e^{-\tau} \frac{\tau^2}{2} h^{++}_2 (\sigma) , \;\; R_{OPE}^{(3)+-}=\cos \phi \; e^{-\tau}  \tau^2  h^{+-}_2 (\sigma) ,
\eeqn
we obtain
\beqn\label{R3PM}
&&R_{OPE}^{(3)+-} \Longrightarrow \frac{i \pi}{12}  \cos  (\phi_2-\phi_3)  |w| \left(-6 \ln  (1-u_1)+\pi ^2-12\right) (\ln  (1-u_1)+2 \ln  |w|-2 \ln 2)^2
\nonumber
\\
&&
\simeq -i2  \pi \cos  (\phi_2-\phi_3)   |w| \ln  (1-u_1) \ln  ^2 |w|+\frac{i \pi ^3}{3}   \cos  (\phi_2-\phi_3) |w| \ln  ^2|w|
\\
 &&-i4  \pi \cos  (\phi_2-\phi_3)  |w| \ln  ^2|w| \nonumber
\eeqn
and
\beqn\label{R3PP}
&& R_{OPE}^{(3)++} \Longrightarrow -i\frac{\pi}{4}    \cos  (\phi_2-\phi_3) |w| \ln ^2(1-u_1)
\\
&& +i\pi   \cos  (\phi_2-\phi_3) |w| \ln 2 \ln (1-u_1)-i\pi   |w| \ln ^2|w|-i\pi   \cos  (\phi_2-\phi_3)
   |w| \ln ^2 2 \nonumber
\\
 && +i2 \pi   \cos  (\phi_2-\phi_3) |w| \ln 2\ln |w| -i\pi   \cos  (\phi_2-\phi_3) |w| \ln (1-u_1) \ln |w| \simeq -i\pi   |w| \ln ^2|w|.\nonumber
\eeqn
Comparing  (\ref{rd3app}), (\ref{R3PP}) and  (\ref{R3PM}) one can see that each power of $\gamma_{1}^{+}(p)$ in the integrand of the remainder function $R_{OPE}$ in (\ref{ROPEexp}) introduces an additional suppression by one power of $\ln(1-u_1)$ in the terms leading in $\ln |w|$.

In this section we found that the BFKL result in the Double Leading Logarithmic Approximation~(DLLA) can be reproduced taking into account only $\gamma_1^{-}(p)$ in the OPE remainder function~(\ref{ROPEexp}). Each power of $\gamma_1^{+}(p)$ introduce an additional suppression in $\ln(1-u_1)$ and the corresponding contributions are not captured by the LLA BFKL analysis.

\end{document}